\documentclass[a4paper,12pt]{article}

\usepackage{amssymb,amsmath,mathbbol,mathrsfs}
\usepackage[usenames,dvipsnames]{color}
\usepackage{hyperref}
\usepackage{xcolor}
\usepackage{stmaryrd}
\usepackage{authblk}
\usepackage{framed}
\usepackage{empheq}
\usepackage{slashed}
\usepackage{hyphenat}
\usepackage{cite}

\usepackage{marginnote}    

\usepackage[left=.55in,right=.55in,top=.8in,bottom=.7in]{geometry}                  % For good copy.

\usepackage{sectsty}
\allsectionsfont{\sffamily\mdseries\upshape} % (See the fntguide.pdf for font help)
\usepackage{tocloft}

\makeatletter
\renewenvironment{abstract}{%
    \if@twocolumn
      \section*{\abstractname}%
    \else %% <- here I've removed \small
      \begin{center}%
        {\bfseries\sffamily\abstractname\vspace{\z@}}%  %% <- here I've added \Large
      \end{center}%
      \quotation
    \fi}
    {\if@twocolumn\else\endquotation\fi}
\makeatother

5

\hypersetup{
	colorlinks=true,         
	linkcolor=MidnightBlue,          
	citecolor=BrickRed,        
	urlcolor=MidnightBlue            
}

\newcommand{\BIBand}{and}%{\&}

\newcommand{\be}{\begin{equation}}
\newcommand{\ee}{\end{equation}}

\newcommand{\Lie}{\mathrm{Lie}}

\newcommand{\F}{{\Phi}}
\renewcommand{\d}{{\mathrm{d}}}
\newcommand{\D}{{\mathrm{D}}}

\newcommand{\G}{{\mathcal{G}}}
\newcommand{\Ad}{{\mathrm{Ad}}}
\newcommand{\pp}{{\partial}}

\newcommand{\fG}{{\mathrm{Lie}(\G)}}

\renewcommand{\bar}{\overline}
\newcommand{\dd}{{\mathbb{d}}}

\renewcommand{\hat}{\widehat}

\newcommand{\lbr}{\llbracket}
\newcommand{\rbr}{\rrbracket}

\newcommand{\cint}{{\int\kern-.87em{<}}}
\newcommand{\sint}{{\int\kern-.75em{\sim}}}
\newcommand{\fint}{{\int\kern-1.00em{\int}}}

\newcommand{\bb}{\mathbb}

\newcommand{\tr}{\mathrm{Tr}}

\renewcommand{\#}{\sharp}
\let\oldmarginpar\marginpar
\renewcommand\marginpar[1]{\oldmarginpar{\color{red}\raggedright\footnotesize #1}}

\title{\sffamily  Notes on a few quasilocal properties of Yang-Mills theory}
\author[1]{\sffamily Henrique Gomes\thanks{gomes.ha@gmail.com}}
\author[2]{\sffamily Aldo Riello\thanks{ariello@perimeterinstitute.ca}}
\affil[1]{\small Trinity College, Cambridge University, Cambridge CB2 1TQ, England}
\affil[2]{\small Perimeter Institute for Theoretical Physics, 31 Caroline St. N., Waterloo, ON N2L2Y5, Canada}

\begin{document}
\maketitle

\abstract{ 

Gauge theories possess non-local features that, in the presence of boundaries, inevitably lead to subtleties. 
In this article we continue our study of a unified solution based on a geometric tool operating on field-space: a connection form. 
We specialize to the $D+1$ formulation of Yang-Mills theories on configuration space, and we precisely characterize the gluing of the Yang-Mills field across regions.
In the $D+1$ formalism, the connection-form splits the electric degrees of freedom into their pure-radiative and Coulombic components, rendering the latter  as conjugate to the pure-gauge part of the gauge potential.
Regarding gluing, we obtain a characterization for topologically simple regions through closed formulas.  These formulas exploit the properties of a generalized Dirichlet-to-Neumann operator defined at the gluing surface; through them, we find only the radiative components and the local charges are relevant for gluing. 
Finally, we study  the gluing into topologically non-trivial regions in 1+1 dimensions. We find that  in this case the regional radiative modes do not fully determine the global radiative mode  (Aharonov-Bohm phases). For the global mode takes a new contribution from the kernel of the gluing formula, a kernel which is associated to  non-trivial cohomological cycles. 
In no circumnstances do we find a need for postulating new local degrees of freedom at boundaries. \\

{\it \noindent
\textbf{Comment:} The partial results of these notes have been completed and substantially clarified in a more recent, comprehensive article from October 2019. (titled: ``The quasilocal degrees of freedom of Yang-Mills theory").
}

}

%%%%%%%%%%%%%%%%%%%%%%%%%%%%%%%%%%%%%%%%%%%%%
% TABLE OF CONTENTS

\newpage
{\hypersetup{	linkcolor=black, }\tableofcontents}

\begin{center}
	\rule{8cm}{0.4pt}
\end{center}
%
%{\hypersetup{	linkcolor=MidnightBlue }}

%%%%%%%%%%%%%%%%%%%%%%%%%%%%%%%%%%%%%%%%%%%%%

%==========================================
\section{Introduction}

In this note, we address the problem of defining quasilocal degrees of freedom (dofs) in electromagnetism and Yang-Mills (YM) theories.
Here, ``quasilocal'' means ``confined in a finite and bounded region'', with a certain degree of nonlocality allowed {\it within} the region. We will call such properties \textit{regional}.
In fact, gauge theoretical dofs cannot be completely localized, since gauge invariant quantities have a certain degree of nonlocality; the prototypical example being a Wilson line.

In electromagnetism, or any Abelian YM theory, although the field strength $F_{\mu\nu}=\pp_{\mu} A_{\nu} - \pp_\nu A_\mu$ provides a local gauge invariant  observable, a  canonical formulation unveils the underlying nonlocality. 
The  components of $F_{\mu\nu}$ do not provide gauge invariant {\it canonical} coordinates on field space: in 3 space dimensions, $\{ E^i(x) , B^j(y) \} = \epsilon^{ijk}\pp_k\delta(x,y)$ is not a canonical Poisson bracket and the presence of the derivative on the right-hand-side is the signature of a nonlocal behavior.

This nonlocality also prevents the factorizability of gauge-invariant observables and  of physical degrees of freedom across regions. This is a viewpoint often adopted in the literature \cite{Polikarpov, Casini_gauge, GiddingsDonnelly}.

From a canonical perspective, the  constraint whose Poisson bracket generates  gauge transformations, namely the Gauss constraint: $G=\pp_iE^i $, is  responsible for these non-local attributes.
The Gauss constraint is an {\it elliptic} equation that initial data on a Cauchy surface $\Sigma$ must satisfy. 
In other words, the initial values of the gauge potential $A_i$ and its momentum $E^i$ cannot be freely specified throughout $\Sigma$. Ultimately, this is the source of both the nonlocality and the difficulty of identifying freely specifiable initial data -- the ``true'' degrees of freedom.  

To summarize, the identification of the quasilocal dofs requires: (\textit{i}) dealing with the Gauss constraint  for the electric field and (\textit{ii}) the separation of pure gauge and physical components of the  gauge potential.
Familiar from the theory of symplectic reductions \cite{AbrahamMarsden}, these two tasks are related but distinct.
 
We will start with the second task, by reviewing a particular gauge invariant formulation of symplectic geometry.
  In \cite{GomesRiello2016, GomesRiello2018, GomesHopfRiello, GomesStudies, RielloSoft}, the present authors  proposed and developed a gauge-invariant formulation of symplectic geometry for electromagnetism and YM theories  that takes advantage of a natural geometric tool on the field-space of these theories: a functional connection-form.  
At the covariant, perturbative level,  this connection-form fulfills precisely the aim of (\textit{ii}): it separates gauge and physical components of perturbations of the gauge field. 

One aim of this paper is  to clarify the mathematics of the functional connection framework in {\it configuration space}, proving its strengths in addressing the interplay of gauge and locality in a dynamical context (previous work was done in the Euclidean context). After introducing all the required machinery and adapting it to the $D$+1 context in section \ref{sec:field_space}, this will lead us automatically to the questions of the Gauss constraint, and the overall effect the split has on symplectic geometric treatments of dynamics. These matters are treated in section \ref {sec:thetaH}. We will show that the connection-form splits the standard symplectic potential into a pure Coulombic and a pure radiative contribution. 

Interestingly, the radiative contributions to the symplectic charges always vanish in the absence of matter. We then add charged matter and show that non-trivial symplectic charges arise in its presence, and these charges are precisely associated to `global gauge transformations', or gauge `reducibility parameters'  \cite{Barnich}.  The association of charges  to global reducibility parameters is too strict, and the puzzle of extending this association to encompass merely asymptotic reducibility parameters  was resolved in \cite{RielloSoft}.

Once one has  the separation of pure gauge and physical components in the gauge potential, i.e. requirement (\textit{ii}) above, a natural question we should answer is: when can we compose,  or ``glue'', physical components from different regions? How does the  gluing work, exactly?   In section \ref{sec:gluing}, we will broach the second main topic of this work: gluing.

For reasons made explicit in \cite{GomesHopfRiello}, our formalism is best adapted to  Yang-Mills theory at the level of perturbations of the gauge and matter fields  around a background configuration. We refer to these as ``perturbations'' or, sometimes, ``modes". At this level, given two regions sharing a boundary, we will show that the physical modes of the gauge fields of the regions can be composed into a global physical mode if and only if there are no physical modes intrinsic to the boundary itself. In other words, if and only if the boundary is \textit{fiducial}. In electromagnetism, this means there are no charged currents at the boundary.\footnote{In section \ref{sec:gluing} the notion of `intrinsic boundary modes' will be appropriately defined.}  

To smoothly compose  physical modes on each region into a global physical mode, we generically require non-trivial gauge transformations of each region. When  the fields are composable as per above, we will show, using standard machinery of the theory of partial differential equations in bounded domains, such as the Dirichlet-to-Neumann operator, precisely which gauge transformations are required in each region. These regional gauge transformations are uniquely determined by the regional physical field content -- actually, by their mismatch at the common boundary. 

The mechanism of gluing underscores the difference between separating horizontal modes and gauge-fixing: had we gauge-fixed, gauge-transformations in each region would be unavailable. The gluing of regional physical modes exploits a flexibility which would not have existed had we gauge-fixed the fields in each region.\footnote{A gauge-fixed description is, of course, gauge-invariant, and  therefore immune to further gauge transformations. Nonetheless, many parallels exist between our work and standard gauge-fixings. We discuss this in appendix \ref{sec:dressing}, based on results obtained in \cite{GomesHopfRiello}.}

 This flexibility plays a decisive role in the last part of our paper. There we exemplify the gluing construction in  one spatial dimensions,  gluing physical perturbations on intervals onto a circle. This case involves a manifold with non trivial cohomology, which gives rise to  an ambiguity in the gluing procedure. It is this ambiguity which underlies the existence of ``would-be-gauge'' global or topological modes (Aharonov-Bohm phases).

Since the presence of true boundary charges seems enough to account for any information lacking in the regions,  our geometric constructions intrinsic to the pre-existing configuration space -- and in particular our results on gluing at fiducial boundaries --  challenge the necessity of introducing \textit{genuinely new} boundary degrees of freedom (edge-modes)  in the context of Yang-Mills theories. In other words, our results challenge the introduction of degrees of freedom beyond the original field content and   irrespectively of the nature of the boundary itself (e.g. \cite{DonnellyFreidel, Geiller:2017xad, Speranza:2017gxd, Camps}).\footnote{Camps has an alternative perspective on what constitutes a fiducial boundary \cite{Camps}.}

 In section \ref{sec:conclusions} we conclude with a summary of the results.

%================================================
\section{Configuration space geometry\label{sec:field_space}}

\subsection{ Horizontal splittings in configuration space}
To start, let us introduce some notation and recall some  basic facts,
 which culminate with a discussion of the generic properties of any functional connection-form. The  introduction of the concrete connection-form -- whose properties constitute the focus of this paper -- will wait until section \ref{sec:radiative}.

\begin{figure}[t]
		\begin{center}
			\includegraphics[scale=0.17]{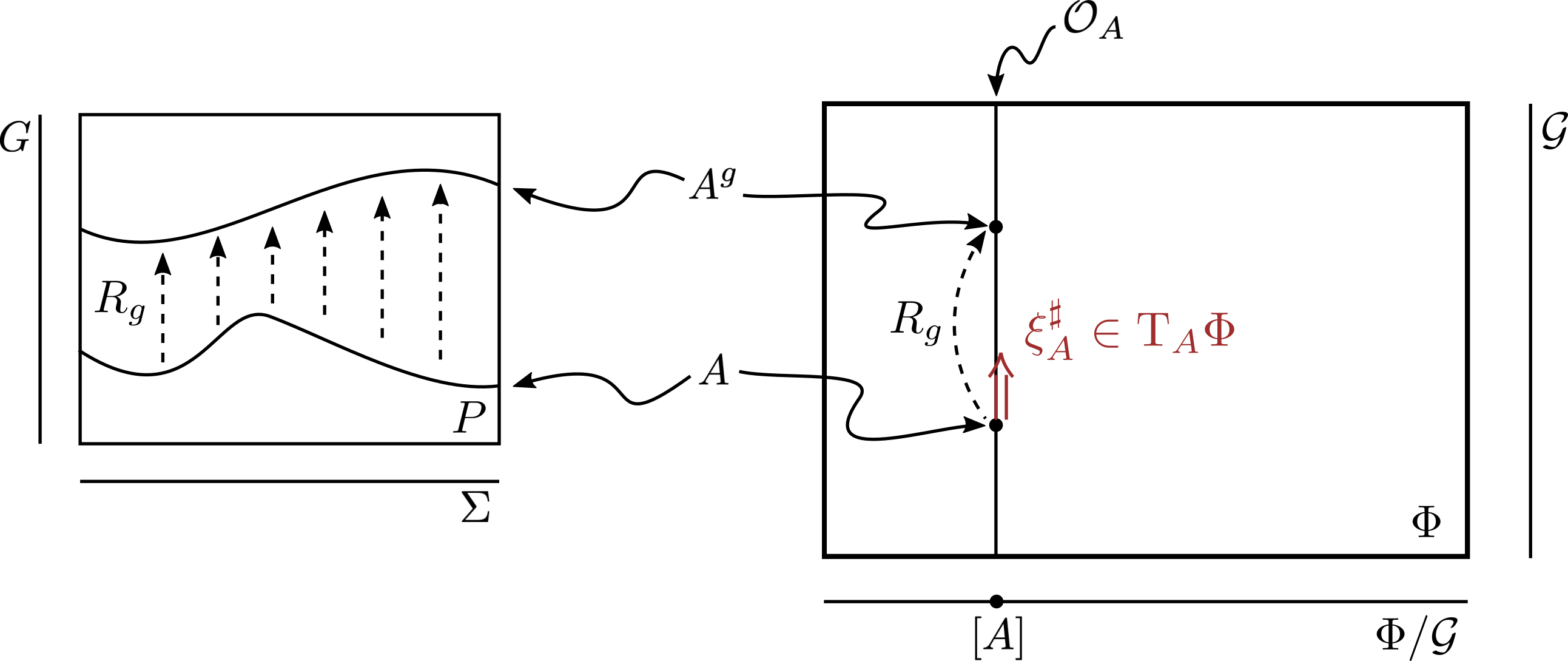}	
			\caption{A pictorial representation of the configuration space $\F$ seen as a principal fibre bundle, on the right. We have highlighted a configuration $A$, its (gauge-transformed) image under the action of $R_g:A\mapsto A^g$, and its orbit $\mathcal O_A \cong \G$. We have also represented the quotient space of `gauge-invariant configurations' $\F/\G$. On the left hand side of the picture, we have ``zoomed into'' two representation of $A$ and $A^g$ as two gauge-related local sections of  a connection $\omega$ on $P$, the principal fibre bundle with structural group $G$ over $\Sigma$. }
			\label{fig1}
		\end{center}
	\end{figure}

 We shall consider a Lagrangian $D$+1 formulation of Yang-Mills theory on a globally hyperbolic spacetime $M$ foliated by equal-time Cauchy surfaces $\Sigma_t$.
 This means that we work in configuration space, that is the space of Lie-algebra valued one-forms, 
 $$ A \in \Phi:=\Omega^1(\Sigma, \Lie(G)).$$ Here, $G$ is a compact semisimple Lie group, called the {\it charge group}.  To distinguish issues of global (topological) nature -- which we will discard in this note with the exception of section \ref{sec:topology} -- from those associated merely with  finite boundaries -- which  constitute our main focus, -- we assume $\Sigma\cong \bb R^D$ and the  configuration space of fields to be given by compactly supported one-forms on $\Sigma$.  This space is denoted by $\Phi_c$. See figure \ref{fig1} for the general structures pertinent to this paper. 

In the following, we will use the notation, $A=A_i\d x^i = A^\alpha_i \d x^i \tau_\alpha$, where  $\{\tau_a\}$ is a basis of  generators of $\Lie(G)$. Notice that  the component of $A$ in the transverse direction to $\Sigma$, $A_0$, is left out of the description, for now. We will reintroduce it in due time.
 
The space of gauge transformations -- i.e. the space of $G$-valued compactly supported functions on $\Sigma$, $\mathcal G:=\mathcal C_c^\infty(\Sigma, G)\ni g$ -- inherits a group structure from $G$ via pointwise multiplication. Call $\G$ the {\it gauge group}.
The fundamental fields $A \in \Phi_c$, or gauge potentials, transform under gauge transformations $g:\Sigma \to G$ as 
\be\label{eq:gt} 
A_i \mapsto A_i^g = g^{-1} A_i g + g^{-1}\pp_i g.
\ee

This formula for the action of a gauge transformation $g$ on a field configuration $A$ provides an action of $\G$ on $\Phi_c$. 
This action induces a fiducial principal fibre bundle structure on configuration space, $\pi:\Phi_c \to \mathcal P$. 
The orbits of this action, $\mathcal O_A$, are called gauge orbits and their space $\mathcal P \cong \Phi_c/\G$ is the space of physical configurations. This is the ``true'', or ``reduced'', or ``physical'',  configuration space of the theory, but it is only defined abstractly through an equivalence relation, and is  most often inaccessible for practical purposes.

An infinitesimal gauge transformation, $\xi\in\Lie(\G)$, defines a vector field tangent to the gauge orbits. 
This will be denoted by $\xi^\#\in \mathrm T \Phi_c$, and its value at $A$ is
\be
\xi_A^\# = \int_\Sigma \d x \; (\D_i\xi)^\alpha(x) \frac{\delta}{\delta A_i^\alpha(x)} \in \mathrm T_A\Phi_c,
\label{eq:hash}
\ee
where $\D_i \xi := \pp_i \xi + [A_i, \xi] $ is the covariant derivative.
The span of the $\xi^\#$'s defines the {\it vertical} subspace of $\mathrm T\Phi_c$, i.e. $V:=\mathrm{Span}(\xi^\#) \subset \mathrm T\Phi_c$. $V$ comprises the ``pure gauge directions'' in $\Phi_c$: vertical changes are ``pure gauge'' and carry no physical information.

The ``physical'' directions in $\mathrm T\Phi_c$ are thus those transverse to $V$, i.e. the {\it horizontal directions} $H\subset\mathrm T\Phi_c$. However, the decomposition $\mathrm T\Phi_c = V \oplus H$ is not canonically defined by the fibre bundle structure. Nevertheless, the {\it choice} of one such decomposition can be encoded in the choice of a connection form on the bundle $\pi:\Phi_c\to\mathcal P$. This is a functional 1-form valued in the Lie algebra of the gauge group,\footnote{Verbally, the functional connection $\varpi$ is usually referred to by its typographical name, \textsc{Var-Pie}.}
\be
\varpi \in \Omega^1(\Phi_c, \Lie(\G)),
\ee 
which is characterized by the following two properties
\be\label{eq:varpi_def}
\begin{cases}
\bb i_{\xi^\#}\varpi = \xi\\
\bb L_{\xi^\#} \varpi = [\varpi, \xi] + \dd \xi
\end{cases}.
\ee
Hereafter, double-struck symbols refer to geometrical objects and operations in configuration space: $\dd$ is the (formal)  field-space deRahm differential,\footnote{We prefer this notation to the more common $\delta$, because the latter is often used to indicate vectors as well as forms, hence creating possible confusions.} $\bb i$ is the inclusion operator of  field-space vectors into  field-space forms, and $\mathbb L_\bb X$ is the  field-space Lie derivative along $\bb X\in\mathfrak X^1(\Phi_c)$. Its action on  field-space forms is given by Cartan's formula, $\bb L_{\bb X} = \bb i_{\bb X} \dd + \dd \bb i_{\bb X} $.

\begin{figure}
		\begin{center}
			\includegraphics[scale=.17]{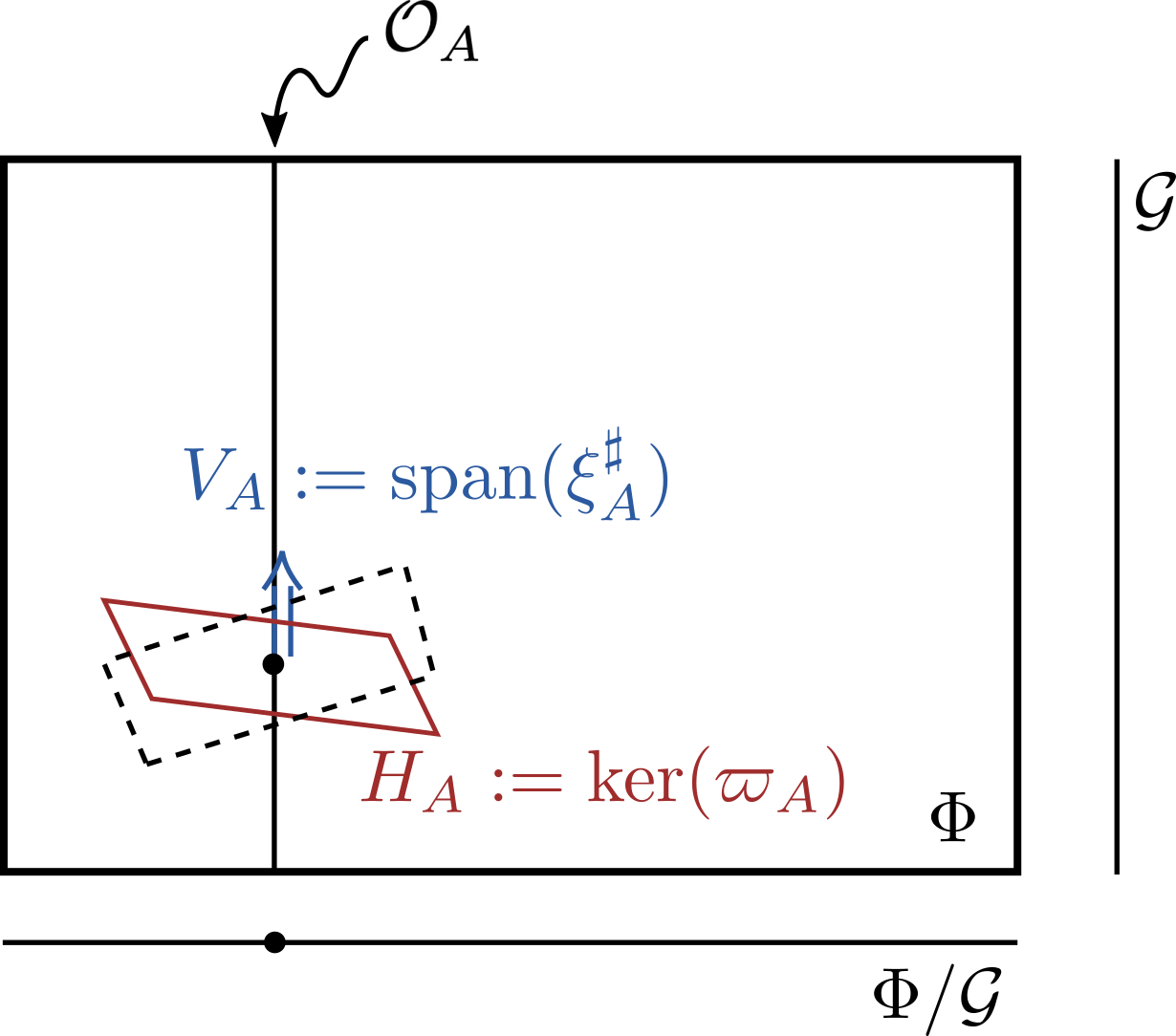}
			\caption{A pictorial representation of the split of ${\rm T}_A \F$ into a vertical subspace $V_A$ spanned by $\{\xi_A^\#, \xi\in\fG\}$ and its horizontal complement $H_A$ defined as the kernel at $A$ of a functional connection $\varpi$. With dotted lines, we represent a different choice of horizontal complement associated to a different choice of $\varpi$.}
			\label{fig2}
		\end{center}
	\end{figure}

The first of the properties \eqref{eq:varpi_def}, the projection property, is required for the definition of a horizontal complement  $H$ to the fixed vertical space $V$ through 
\be
H := \ker \varpi.
\ee
The second of the properties \eqref{eq:varpi_def}, on the other hand, ensures the compatibility of the above definition  with the group action of $\G$ on $\Phi_c$, i.e. it embodies covariance under gauge transformations. See figure \ref{fig2} for a pictorial representation. 

The term $\dd\xi$ on the right hand side of the covariance condition is only present if $\xi$ is chosen differently at different points of $\Phi$, i.e. if $\xi$ is an infinitesimal {\it field-dependent} gauge transformation.\footnote{The form of this equation can be deduced from the standard transformation property, $\bb L_{\xi^\#} \varpi = [\varpi, \xi]$, for $\xi$'s constant throughout $\Phi$ and the projection property of $\varpi$ which holds pointwise in  field-space. See \cite{GomesHopfRiello}.} Gauge-fixings are a typical example of field-dependent gauge transformations. 

Field dependent gauge transformations have applications to gravity,\footnote{E.g., in a 3+1 formulation of general relativity, the shift needs to accommodate field-dependence to encode the commutation algebra \cite{HKT}.}  and must be contemplated in the presence of gauge-fixings or changes of gauge. But they  are more easily justified by two simple observations. Firstly, their introduction does not require any extra geometrical information while providing, as we shall discuss, a refined diagnostic tool in the presence of boundaries (in their absence, generalizing to field-dependent gauge transformations changes nothing). Secondly, and  more abstractly, if gauge has to be only descriptive fluff, different configurations should be allowed to be acted upon by distinct and independent gauge transformations. Later on, we will provide a further, more concrete  justification, which is specific to the case with boundaries. It is there that the introduction of $\dd \xi\neq 0$ pays dividends.
  
Now, given a connection-form characterized by \eqref{eq:varpi_def},   alongside $\dd$ we can introduce the horizontal differential, $\dd_H$. Horizontal differentials are by definition transverse to the vertical, pure gauge, directions: $\bb i_{\xi^\#} \dd_H A= 0$. We opted to emphasize the geometrical properties of $\varpi$ by using the subindex $H$ (meaning, horizontal). However, we could have as well opted for the alternative notation $\dd_H \equiv \dd_\varpi$, more explicit about the choice of connection and reminiscent of the notion of ``covariant derivative''.
Indeed, the horizontal differentials of $A_i$ and $E^i$  (seen as coordinate functions on $\Phi_c$) are given by  the following ``covariant differentials"
\be\label{eq:dH}
\dd_H A_i = \dd A_i - \D_i \varpi
\qquad\text{and}\qquad
\dd_H E^i = \dd E^i - [E^i, \varpi].
\ee
 The covariance property of these expressions, i.e. the homogeneous action of (possibly field-dependent)  gauge transformations on  the horizontal differentials,  is easy to show, and reads:
\be
\bb L_{\xi^\#} \dd_H A_i = [\xi, \dd_H A_i] 
\qquad\text{and}\qquad
\bb L_{\xi^\#} \dd_H E^i = [\xi, \dd_H E^i] .
\ee
 For details see \cite{GomesRiello2016}.

%============================
\subsection{Horizontality in the $D$+1 decomposition }

We now further dissect the properties of the electric field in relation to horizontality  in configuration space. 

Since we are working on configuration space, the electric field $E$ is here understood as a functional of $A$ (and $A_0$ -- to which we will come back shortly), through  the usual formula $E_i := n^\mu F_{\mu i}[A]$, where $n^\mu$ is the unit timelike normal to $\Sigma$.

For simplicity, we will assume that $\Sigma$ belongs to an ``Eulerian" foliation of spacetime, i.e. to a foliation whose lapse is equal to one and whose shift vanishes.\footnote{ If the particular spacetime does not globally admit such a foliation, we restrict our attention to a patch of it that does, which always exists.} In other words, $\Sigma$ is an equal-time hypersurface in a spacetime with metric $\d s^2 = -\d t^2 + g_{ij}(t,x)\d x^i \d x^j$.

 Including nontrivial lapse and shift is straightforward, but makes some formulas more cluttered, without adding anything to our present arguments. However, see \cite{RielloSoft} for a situation where the introduction of a nontrivial shift plays a crucial role in dealing with asymptotic gauge transformations and charges.

With this simplifying hypothesis of Eulerian foliations,  the electric field is
\be
E_i = \dot A_i - \D_i A_0 ,
\label{eq:Efield}
\ee
the dot standing for the derivation with respect to $x^0=t$.
Three dimensional indices $i,j,\dots$ will be lowered and raised with the induced metric on $\Sigma$, $g_{ij}$, and its inverse, $g^{ij}$. 
Also, the derivative $\D_i$ is here understood to be both gauge- and space-covariant, i.e. symbolically  $\D_i = \nabla^{\text{LC}}_i + A_i $, where LC indicates the Levi-Civita connection of $g_{ij}$.

Now, recall that $A_0$ is non-dynamical and plays the role of the Lagrange multiplier enforcing the Gauss constraint, $\D_i E^i \approx 0$. As such, in the configuration space treatment, $A_0$ is often gauge-fixed to zero.

However, $A_0$ fulfills another, perhaps less appreciated role, as the enforcer of gauge invariance of $F_{\mu\nu}$ (and hence $E^i$) under {\it time-dependent} gauge transformations. 

If $A_0$ is gauge-fixed to be some function (e.g. a constant function), thereby  relinquishing its own gauge-transformation properties, the composite field loses the corresponding invariance, and therefore one must also  exclude time-dependent gauge transformations from consideration.    

 Thanks to the connection form, we can obviate this problem by defining
\be
\label{eq:A0}
A_0 = \varphi + \varpi(\bb v).
\ee
Here, $\bb v$ is the velocity vector in configuration space, i.e. the tangent vector to the curve $A_i(t)$ defining the evolution of $A_i$ in $\Phi_c$:
\be
\bb v_A = \int \d x\, \dot A^\alpha_i(x) \frac{\delta }{\delta A^\alpha_i(x)} \in\mathrm T_A\Phi_c.
\label{eq:v}
\ee
Also, in \eqref{eq:A0}, we used the notation $\varpi(\bb v )\equiv \bb i_{\bb v } \varpi$ for mere typographical reasons.

Notice that, with this definition,  $A_0$ depends on the history of the configuration variable $A_i$, while $\varphi$ is a new variable independent of $A_i$.

In electromagnetism, definition \eqref{eq:A0} implies that $\varphi$  must not transform under gauge transformations, since the entire burden of ensuring the correct  transformation properties of $A_0$ is carried by the configuration space variable $A_i$. Indeed, consider a time-dependent gauge transformation $\xi$, $\dot \xi\neq 0$. Since $\delta_\xi A_i = \pp_i\xi$, one has from \eqref{eq:varpi_def} and \eqref{eq:hash}
\be
\label{eq:A0_transfEM} 
\delta_\xi \dot A_i = \pp_i \dot \xi, \qquad \delta_\xi A_0 = \varpi(\dot\xi^\#) = \dot \xi \qquad\text{(Abelian)}.
\ee
In non-Abelian Yang-Mills theory, the burden of the inhomogeneous transformation of $A_0$ is still carried by $A_i$ (and $\varpi$), so that  $\varphi$ must simply transform covariantly in the adjoint representation, $\delta_\xi \varphi =[\varphi, \xi]$.
The analogues of \eqref{eq:A0_transfEM} once again follow directly from \eqref{eq:A0}, \eqref{eq:v} and \eqref{eq:varpi_def}, but the proof is considerably more subtle (and interesting): it makes full use of the geometrical framework we are advocating for and does justice to the role played by $\varpi$ and the field-dependent gauge transformations. The reader can find this proof in appendix \ref{app:A0gauge}. 

Now, with the above definition of $A_0$,  the electric field \eqref{eq:Efield} reads
\be
\label{eq:E}
E_i = \dot A_i - \D_i A_0 =\dot A_i^H - \D_i \varphi,
\ee
where we introduced the horizontal component of $\dot A$, i.e. $\dot A^H := \bb i_{\bb v}( \dd_H A) = \dot A - \D_i \varpi(\bb v)$ and used equation \eqref{eq:A0}.
Of course, from the transformation properties of $A_i$ and $A_0$, it follows that $\delta_\xi E_i = [E_i , \xi]$, as it should (again, we refer to appendix \ref{app:A0gauge} for a proof in the non-Abelian case).
Notice that equation \eqref{eq:E}  emphasizes the separation in $E_i$ of the {\it horizontal} dynamics of $A_i$ from the contribution due to $\varphi$.

%==========================================

%====================
\section{Horizontality and symplectic geometry \label{sec:thetaH}}

\subsection{Symplectic structure in finite regions and horizontality}\label{sec:symp_finite}

In this section, we introduce the promised gauge-invariant symplectic potential\footnote{To keep the language lighter, we won't distinguish between symplectic and {\it pre}symplectic potential or form.  } over finite regions.
This corresponds to a horizontal version of the standard symplectic potential.

We start with the standard symplectic potential, $\theta \in \Omega^1(\Phi)$, which reads 
\be\label{eq:symp_pot}
\theta = \int_\Sigma \sqrt{g} \,\tr( E^i \dd A_i).
\ee
Here, the integration takes place over the entire, unbounded, Cauchy surface $\Sigma$.
It is manifestly gauge invariant under field-independent gauge transformations, $\dd \xi = 0$. 
From $\theta$, and for field-independent gauge transformations, one can construct the Noether charges:\footnote{This formula, often written as $Q[\xi] := \theta(\delta_\xi)$,  is not general for {\it any} gauge theory, but applies in this form to Yang-Mills theories, whose Lagrangian {\it density} is fully invariant  (and not just up to total derivative) under gauge transformations,  and with the choice of the standard symplectic potential of equation \eqref{eq:symp_pot}.}
\be\label{eq:Q_theta}
Q[\xi]:=\bb i_{\xi^\#}\theta.
\ee
These charges are the symplectic generators corresponding to the gauge transformation $\xi$,
\be\label{eq:Ham_flow}
\bb i_{\xi^\#} \Omega = - \dd Q[\xi] \qquad (\dd \xi = 0), 
\ee
where $\Omega :=\dd \theta$ is the symplectic 2-form.

This fundamental equation, \eqref{eq:Ham_flow},  is essentially a re-writing, through Cartan's formula $\bb L_{\bb X} = \bb i_{\bb X}\dd + \dd \bb i_{\bb X}$, $\bb X = \xi^\#$, of the gauge invariance of $\theta$. In other words, the gauge invariant condition $\bb L_{\xi^\#} \theta = 0$ translates into the Hamiltonian nature of the gauge flow $\xi^\#$, which is generated by $Q[\xi]$ as given in \eqref{eq:Q_theta}.

 In fact we can extend this invariance. For  field-{\it dependent} gauge transformations, $\dd \xi \neq 0$, the violation of the gauge invariance of $\theta$ is proportional to the Gauss constraint, $\D_i E^i \approx 0$.
Therefore, {\it on-shell} ($\approx$) of the Gauss constraint, we have
\be
\bb L_{\xi^\#}\theta=\bb i_{\xi^\#} \Omega + \dd Q[\xi] = \int_\Sigma \sqrt{g}\,\tr( E^i \D_i \dd \xi) \approx 0.
\ee
This means that, although the flow equation \eqref{eq:Ham_flow} does not strictly hold, its violations  do not ``interfere" with the gauge structure, since they vanish on-shell of the Gauss constraint.\footnote{ It is easy to translate this example to standard Hamiltonian formalism: there, for a field-dependent gauge-parameter -- i.e. a smearing  $\lambda^\alpha(p,q)$ of first class constraints $H_\alpha$ -- we obtain $$\{H_\alpha(\lambda^\alpha(p,q)), F(p,q)\}=\{H_\alpha, F(p,q)\}(\lambda^\alpha(p,q))+H_\alpha(\{\lambda^\alpha(p,q), F(p,q)\})\approx \{H_\alpha, F(p,q)\}(\lambda^\alpha(p,q))$$ I.e. the constraints still effect a homomorphism from the gauge group of transformations into canonical transformations in phase space.  }

This fact is closely related to the vanishing of the Noether charges $Q[\xi]$, in the absence of boundaries, and on-shell of the Gauss constraint:
\be
Q[\xi] \approx 0.
\ee 

In the absence of boundaries,  it's relatively uncontroversial that there really \textit{should} be no difference in considering $\dd \xi=0$ or $\dd\xi\neq 0$.  However, as we will see below,  in the presence of boundaries, there \textit{is} a difference between $\dd \xi=0$ and $\dd \xi\neq 0$, and it shows up uncontrovertibly in the symplectic formalism.

Usually,  the presence of boundaries implies restrictions on both the field content and the gauge group, i.e. both $\xi$ and $A$ obey rigid boundary conditions (see e.g. \cite{Giulini_gauge} for a discussion). 
But,  by not restricting our primary fields, we are treating our boundaries as \textit{fiducial}.\footnote{ By fiducial here we mean: the boundaries do not impose any further restrictions on the field. In section \ref{sec:gluing}, we will amend this definition to also encompass the fact that they can be ``erased''  by gluing. In that context, a mathematical definition of ``fiducial'' will be put forward.} In that case  there is no reason to treat bulk and boundary gauge transformations any differently.
And full  commitment to the idea that ``gauge degrees of freedom are unphysical or redundant'' obliges us to consider also the possibility that gauge transformations are contingent on the field configuration, $\dd \xi\neq 0$.\footnote{One should also note that although for principal fibre bundles one has a natural action of the charge group, $G$, this is not the case for associated bundles. There, outside of the normal subgroup of $G$,  one cannot define a ``constant''  action of the gauge group: it can only be constant with respect to a given section (see \cite{kobayashivol1} or \cite{Giulini_diff}). Therefore, there is already some reference to the particular representation of $[A]$. Although this serves as a good motivation to encompass $\dd \xi\neq 0$, it is not clear to us if this dependence on the section is either necessary or sufficient for this purpose. }  
 
 To see the difference $\dd \xi\neq 0$ incurs in the symplectic formalism, we first note that the definition of $\theta$ is readily applicable if we replace $\Sigma$ with $R\subset \Sigma$, for $\pp R\neq\emptyset$. 
This regional version of $\theta$ morally captures the regional degrees of freedom present in $R\subset \Sigma$.
However, the properties of the resulting symplectic potential $\theta_R$ are dramatically different: $\theta_R$ fails to be fully gauge invariant (under field-dependent gauge transformations) and the associated Noether charges are non-vanishing, even on-shell of the Gauss constraints.
This failure is due to the implicit {\it integration by parts} required for those results. 
Instead, regionally, we obtain
\be\label{eq:non_inv_theta}
\bb L_{\xi^\#}\theta_R \approx \oint_{\pp R} \sqrt{h}\, \tr( E_s \dd \xi)
 \qquad\text{and}\qquad\bb i_{\xi^\#}\theta_R \approx \oint_{\pp R} \sqrt{h}\, \tr( E_s \xi),
\ee  
where the subscript $s$ denotes contraction with the unit outgoing normal $s^i$ to $\pp R \subset \Sigma$, and $h$ denotes the determinant of the induced metric on $\pp R$, $h_{ij}=\imath_{\pp R}^*g_{ij}$ (for the embedding $\imath_{\pp R}$ of the boundary into $R$).

 These results, illustrated by equations \eqref{eq:non_inv_theta},  can be interpreted either as a sign that ``gauge transformations turn physical at boundaries'',   and this is why the variation $\dd \xi$ makes its appearance, or as indicating that, although gauge is still unphysical fluff, the wrong geometric tools are being used. 
 
   If the former is the case, there are further mysteries to solve: e.g. why is it that these new dynamical properties appear only a boundaries? What can this mean physically?\footnote{ For a full criticism, see \cite{GomesHopfRiello, RielloSoft}. One can try to make sense of charges with these degrees of freedom \cite{DonnellyFreidel}, but charges then become independent of the original field-content  in the bulk of the region, an undesirable property in our point of view.}
  
 Here, instead, we embrace the latter viewpoint: as anticipated, the non-invariance of $\theta$ is resolved, and the discrepancy  between  gauge charges for regions with and without boundaries is erased, by introducing the regional {\it horizontal symplectic potential}
\be
\theta^H_R := \int_R \sqrt{g} \, \tr( E^i \dd_H A_i). 
\label{eq:thetaH}
\ee
 In other words, we claim that the horizontal symplectic potential gives a unified treatment of bounded and unbounded regions. 
Indeed, from \eqref{eq:dH}, it is straightforward to recognize that the regional horizontal symplectic potential is gauge invariant and produces vanishing gauge Noether charges:
\be
\bb L_{\xi^\#}\theta^H_R  \equiv 0
\qquad\text{and}\qquad
\bb i_{\xi^\#}\theta^H_R   \equiv 0.
\label{eq:thetaHzero}
\ee
Easy algebraic manipulations verify that the associated symplectic form is horizontal and closed $\Omega^H_R := \dd_H \theta_R^H = \dd \theta^H_R$, and, in cases of interest, differs from the standard symplectic form by an exact form (in  configuration space and spacetime), $\Omega^H_R \approx \Omega_R+\dd \d \alpha$ (where $\approx$ denotes use of the Gauss law) \cite{GomesRiello2016, GomesRiello2018, GomesHopfRiello}.
 On a compact Cauchy surface  without boundaries, or on an infinite region, but taking into account only compactly supported fields, there is therefore no difference between using the standard or the horizontal symplectic form.

%====================
\subsection{Radiative degrees of freedom and the Singer-DeWitt functional connection }\label{sec:radiative}

An important question which naturally arises is: what kind of degrees of freedom do actually appear in $\theta_R^H$ and what is their relation to the nonlocality implicit in  the Gauss constraint?

To answer this question, we will need to  go beyond  the connection-form's general properties used above, and introduce a specific connection-form.  Since from the mere fibre bundle perspective, there is no canonical choice of $\varpi$, we need a new ingredient -- the kinematical supermetric $\bb G$ -- which will pick out a preferred choice. To fully justify our choice, we first need to further examine the meshing of the electric field $E$ with the  configuration-space geometry we have introduced.

To start with, we insert equation  \eqref{eq:E} for $E$ into the definition of $\theta^H_R$ in \eqref{eq:thetaH}, giving
\be
\theta^H_R = \int_R \sqrt{g} \, g^{ij} \tr\Big((\dot A^H_i - \D_i \varphi )\dd_H A_j \Big).
\label{eq_thetaH}
\ee

This formula requires two comments.
First, whereas the first term represents what we might have intuitively expected -- the pairing of a gauge-invariant (horizontal) degree of freedom, $\dd_H A_i$, with its corresponding horizontally-projected velocity, $\dot A^H_i$ -- the term containing $\varphi$ looks superfluous: what is its role? Second, as we mentioned in the introductory remarks, and have since forgotten: $E$ is not freely specifiable, as it must satisfy the Gauss constraint.

It turns out that both  problems can be solved with one geometrical stroke. 
Fortuitously, the tool required for this resolution is the simplest explicitly geometrical connection-form in configuration space. 

The kinetic term in the Lagrangian, $K$, is quadratic in the velocities $\dot A$. 
Geometrically, it can be written as the square of the velocity $\bb v$, i.e. $K = \frac12\bb G(\bb v, \bb v)$, where
\be
\bb G(\bb X, \bb Y) = \int_R \sqrt{g} \, g^{ij} \tr (\bb X_i \bb Y_j)
\qquad\text{for any $\bb X, \bb Y\in\mathrm T\Phi_c$}.
\ee

Of course, this supermetric defines a conversion of velocities (vectors) into momenta (covectors) and hence also harbors  the definition of the symplectic potential.   
To illustrate these properties, we  first introduce the following vector on configuration-space 
\be
\bb E:= \int_R  E_i^\alpha(x) \frac{\delta}{\delta A_i^\alpha (x)} = \hat H(\bb v) - \varphi^\#,
\label{eq:numero22}
\ee
where $\hat H(\bb v) := \bb v - \varpi^\#(\bb v)$ is the horizontal projection of $\bb v$, and where in the second equality we used equation \eqref{eq:E} for the electric field $E_i$ as well as the definition of the $\cdot^\#$ operator \eqref{eq:hash} (this sends function(al)s valued in $\Lie(\G)$ into vertical vector fields on $\Phi_c$).
The vector field $\bb E$ can be thought of as a modified velocity vector $\bb v$, better adapted to describe the physics.   It is important to notice that, although $\varphi^\#$ is vertical in $\Phi_c$, it is not pure gauge: it is a physically measurable component of the electric field. The  relation between $\varphi^\#$ and gauge degrees of freedom will be clarified soon.

Using the kinematical supermetric $\bb G$, we can now  turn $\bb E$ into a  1-form on configuration space: this is precisely the standard symplectic potential $\theta= \bb G( \bb E, \cdot)$ given in \eqref{eq:symp_pot}. Similarly, the horizontal symplectic potential of \eqref{eq:thetaH} can be written as
\be
\theta^H = \bb G(\bb E, \hat H(\cdot )) = \bb G (\hat H(\bb v) , \hat H(\cdot)) - \bb G (\varphi^\#, \hat H(\cdot)),
\label{eq_thetaH_G}
\ee
where the last equality is just a re-writing of equation \eqref{eq_thetaH}.

We are finally in place to introduce a privileged connection form on configuration space. It is manifest from \eqref{eq_thetaH_G} that $\varphi$ would drop from the horizontal symplectic structure if the notion of horizontality was tied to that of $\bb G$-orthogonality with respect to the fibres;  then any vector tangent to the fibres -- i.e. one of the form $\xi^\#$ -- would be orthogonal to a horizontally projected element.

Denoting with $\perp$ the fact that horizontality is being defined via orthogonality  to the gauge orbits $\mathcal O_A$ relative to the kinematical metric, we then obtain from these considerations:
\be
\theta^{\perp} := \theta^{H_\perp} = \bb G(\bb E, \hat H_{\perp}(\cdot )) = \bb G (\hat H_{\perp}(\bb v) , \hat H_{\perp}(\cdot)) ,
\ee  
which contains {\it no} contribution from $\varphi$.

In gauge theories, the gauge invariance of $\bb G$, in the sense of $\bb L_{\xi^\#}\bb G =0$ \cite{DeWitt_Book}, is manifestly satisfied. This property guarantees that the corresponding metric-induced connection form, named Singer-DeWitt connection (SdW) $\varpi_{\perp}$, satisfies the covariance property in \eqref{eq:varpi_def}.\footnote{A weaker condition is in fact sufficient \cite{GomesHopfRiello}.}

Our choice of naming -- Singer-DeWitt connection -- makes reference to its original introduction, made in the absence of boundaries, by Singer and DeWitt in the 1960s and 1970s \cite{DeWitt_QG1, DeWitt:1967ub,Singer:1981xw, Singer:1978dk}  (see  section 4.3 in \cite{GomesHopfRiello} for a complete reference list). It was  revived, now in the presence of boundaries, in \cite{GomesRiello2016, GomesRiello2018, GomesHopfRiello}, where it was  less transparently denoted as $\varpi_\text{SdW}$.
 In this article, we  provide new and more compelling motivations to make use of this connection form on {\it configuration space}.

Deriving the defining equation for $\varpi_{\perp}$ from $\bb G$ is straightforward; it is equivalent to the imposition of the following requirement  (see references above):
\be
\bb G(\xi^\#, \hat H_\perp(\bb X) ) \equiv \bb G( \xi^\#, \bb X - \varpi^\#(\bb X) ) \stackrel{!}{=} 0
\label{eq:SdWG}
\ee
for all choices of $\xi$ and $\bb X$.

In the absence of boundaries, the resulting equation for $\varpi_{\perp}$ is the elliptic equation
\be\label{eq:varpi_nobdary}
\D^2\varpi_{\perp} = \D^i \dd A_i,
\ee
where $\D^2$ is the gauge covariant version of the Laplace-Beltrami operator ($\D^2 = \Delta$, in electromagnetism).
The SdW connection is therefore nonlocal in space (but local in configuration space!).

Its kernel is the space of ``SdW-horizontal field variations'', which correspond by definition to  configuration-space vectors which are orthogonal to the gauge orbits in $\Phi$. 
As it is easy to see from \eqref{eq:varpi_nobdary}, horizontal variations must therefore be covariantly divergenceless. In electromagnetism they correspond to transverse photons.\footnote{Contract a horizontal perturbation $\bb h$ with equation \eqref{eq:varpi_nobdary} and use $\bb i_{\bb h} \varpi_{\perp} \equiv 0$ and $\bb i_{\bb h} \dd A \equiv \bb h$. This kind of projections is common in QED, $\hat H^i_j= \bb{1}^i_j-\partial_j\Delta^{-2}\partial^i$.}

 However,  the above equations are naturally complemented by appropriate  conditions at the boundaries, when they are present. 
These conditions come from  integration by parts in \eqref{eq:SdWG}, in order to obtain equation \eqref{eq:varpi_nobdary}.
Requiring that also the so obtained boundary term vanishes for all choices of $\xi$ and $\bb X$, readily gives the following elliptic boundary value problem \cite{GomesRiello2018, GomesHopfRiello}:
\be
\begin{cases}
\D^2 \varpi_{\perp} = \D^i \dd A_i & \text{in $R$}\\
\D_s \varpi_{\perp} = \dd A_s & \text{at $\pp R$}
\end{cases}
\label{eq_SdW}
\ee
where as above the subscript $s$ stands for contraction with   $s^i$, the unit normal to $\pp R$.
 In electromagnetism, this boundary value problem for the SdW connection is of the Neumann type, while in non-Abelian YM theories it is of a specific Robin type. 

 It is crucial to stress this point:  the above boundary value problem involves boundary conditions for $\varpi_{\perp}$, but {\it not} for the gauge field $A_i$ nor for its perturbations.   {\it No} boundary conditions are imposed on the actual field content; the region is fiducial, only ``virtually'' carved out of a Cauchy surface $\Sigma$.
The boundary conditions on $\varpi_{\perp}$ ensure that the connection in a region $R$  -- and the corresponding horizontal projections -- are uniquely defined. 
But it does not restrict the value of the fields that we are considering there, neither the gauge potentials nor the gauge transformations, and  neither in $R$ nor at $\pp R$.

Thus, in the presence of boundaries, SdW-horizontal perturbations, i.e. those in the kernel of $\varpi_{\perp}$, are covariantly divergenceless in the bulk and vanish when contracted with $s^i$ at the boundary. In particular, the SdW-horizontal contribution to the electric field vector $\bb E_i$ \eqref{eq:numero22},  i.e. 
\be
\hat H_\perp(\bb v) = \bb v - \varpi_{\perp}^\#(\bb v) =  \int \Big(\dot A - \D \varpi_{\perp}(\bb v)\Big)\frac{\delta}{\delta A} =: \int \dot A^{\perp} \frac{\delta}{\delta A}  ,
\ee
 is such that
 \be
 \begin{cases}\label{eq:sdw_A}
 \D^i \dot A^{\perp}_i=0 & \text{in $R$}\\
 s^i\dot A^{\perp}_i=0 &  \text{at  $\pp R$}.
 \end{cases}
 \ee 
As we will investigate in greater detail in the next section, this means in particular that the SdW-horizontal contribution to $E$  drops from the Gauss law in $R$.
 
Now we can confirm the  properties alleged in the beginning of the section for the SdW connection in relation to the symplectic potential. The explicit formula for the SdW-horizontal symplectic form in $R$ is:\footnote{$\dd_\perp := \dd_{H_\perp}$ is the horizontal differential associated to $\varpi_{\perp}$.}
\be\label{eq:theta_perp}
 \theta^{\perp} =  \int_R  \sqrt{g} \, g^{ij} \dot A^{\perp}_i \,\dd_{\perp} A_j.
\ee
This equation emphasizes the fact that horizontal perturbations are conjugate to horizontal velocities -- an exclusive property of  of the SdW connection. 

As we will see, the disappearance of $\varphi$  from $\theta^\perp$  has an explanation:  through the choice of the SdW connection, $\varphi$ is completely constrained by the matter distribution. In other words,  its initial value cannot be freely specified and it does not have any dynamics of its own apart from that implied by the instantaneous charge distribution. It is, in this sense, a non-dynamical variable, i.e. it does not embody an actual degree of \textit{freedom} of the system.

 We finish this subsection with one important comment.  
The base-point field-dependence of the non-Abelian case makes it clear that we only decompose  field-space vector fields $\bb X \in \mathrm T \Phi_c$, and {\it not} the configurations $A\in\Phi_c$. This distinction is particularly important in the non-Abelian context, where $\varpi_\perp$ is not flat. But in the Abelian context, given a base-configuration, the SdW connection is flat and can therefore be integrated in a path-independent way to associate a ``decomposition'' to field-configurations as well -- which is however best understood in terms of so-called dressings. See  section \ref{sec:dressing}  for a brief and \cite{GomesHopfRiello} for a detailed discussion of this important point. 

Alternatively, one can use the affine structure of $\F_c$ to identify, given an arbitrary reference configuration, any other configuration with a vector. Such a vector can then be decomposed through $\varpi$ at the reference configuration \cite{Babelon:1979wd}.\footnote{This construction can be used to explain the geometrical meaning of the Faddeev-Popov determinant.} This idea, however, is also problematic in the non-Abelian case \cite{Gribov:1977wm, Singer:1978dk} and we feel it clashes with the infinitesimal nature of  symplectic geometry.

%====================
\subsection{Gauss constraint}\label{sec:Gauss}

In the previous section we have used the decomposition of the electric field (into a horizontal velocity $\dot A^\perp$ and a vertical $\varphi^\#$) and a choice of a connection form ($\varpi=\varpi_{\perp}$) to prove that the corresponding SdW-symplectic form depends on the electric field $E$ only through the purely  horizontal component $\dot A^\perp$ of the configuration-space vector $\bb E$.\footnote{For an earlier construction, which is technically very similar to the present one, but different in spirit, see \cite{AliDieter}. See also \cite{Ali2} which comments on the multipole charges related to $\varphi_\perp$ and encoded in the vertical component $\theta_V := \theta - \theta_\perp$ (see equations \eqref{eq:32} and \eqref{eq:decomposition}).}
 
So, the question naturally arises: what is the role of the vertical component of $\bb E$?
 Let us first clarify a point that could be a source of confusion.  With clarification in mind, we momentarily revert to more standard notation: considering a generic perturbation of the gauge potential, $\delta A_i$, its vertical component with respect to a certain connection $\varpi$ constitutes that component of $\delta A_i$ which is interpretable as ``pure gauge''.
Conversely, the horizontal component of $\delta A_i$ is interpreted as encoding the ``physical'' change. 
A  different matter is the vertical/horizontal decomposition of the configuration-space vector $\bb E$ associated to the electric field $E_i$. 
By performing that decomposition, we do {\it not} imply that the vertical part of the electric field $E_i$ is pure gauge and the horizontal one is physical.
 In fact, the decomposition applies equally well to the Abelian and non-Abelian case, and in the Abelian, clearly both $E_i$ and $\varphi$ are gauge invariant quantities.

Instead, by using the SdW-horizontal symplectic form: (\textit{i}) the horizontal part of $\bb E$ identifies the ``radiative'' component of the electric field, whereas (\textit{ii}) its vertical part consists of the Coulombic component of the electric field. This Coulombic component is fully constrained by the Gauss law and appears in $\theta$  conjugated to the pure-gauge part of $\delta A_i$ (with respect to $\varpi_{\perp}$). 
We  now turn  to proving these statements.

Using the SdW connection  $\varpi_\perp$, and denoting by $\varphi_\perp$ the corresponding SdW-vertical component of $\bb E$, it is immediate to see that the burden of the Gauss law is fully entrusted to $\varphi$ (we will generalize to the case with matter in the next section):
\be
0\approx \D^i E_i = \D^i \dot A_i^H - \D^2 \varphi  =  - \D^2 \varphi_\perp ,
\label{eq:32}
\ee
where we specialized to $\varpi=\varpi_{\perp}$ only in the last equality, $\D^i \dot A_i^\perp =0$ (notice that the choice of $\varpi$ alters the corresponding definition of $\varphi$, if $A_0$ is considered as the fundamental object).

Therefore, the SdW connection is not only the {\it unique} functional connection with no contribution of $\varphi$  in $\theta^H$, but it also identifies $\varphi_\perp$ with those components of $E$ which are fully  and uniquely determined by the Gauss constraint and are, in \textit{that} sense, non-dynamical.

One can recover the standard form of the symplectic potential, by reintroducing the vertical component of $\dd A$, i.e. $\dd A=\dd_H A+\D\varpi$,  to \eqref{eq:theta_perp}, obtaining the two alternative expressions:
\begin{align}
\theta & = \theta^{\perp} + \int_R \sqrt{g} \,\tr( E^i \D_i\varpi_{\perp})  \approx \theta^\perp + \oint_{\partial R} \sqrt{h} \, \tr(E_s  \varpi_{\perp} ) \notag\\
& = \theta^\perp + \int_R \sqrt{g} \, \tr(\D^i \varphi_\perp \D_i \varpi_{\perp} ) \approx \theta^\perp - \oint_{\partial R} \sqrt{h} \, \tr( \D_s \varphi_\perp \, \varpi_{\perp} ).
\label{eq:decomposition}
\end{align}
The vertical contribution to $\theta$ is the instantaneous ``Coulombic'' contribution to the symplectic form which, on-shell of the Gauss law, results in a pure boundary term. 
This is the only contribution that depends on $\varphi_\perp$, which we have therefore isolated.  The normal component of the electric field at the boundary, the ``Coulombic term'', is thus identified with the  vertical part of $E$.

%========================================

 In the next  section, we will first extend the results  of section \ref{sec:Gauss} to the presence of matter and then specialize them to SdW-horizontality.
In doing so, we will see that certain natural notions of non-vanishing global charges manage to survive \cite{GomesRiello2018,GomesHopfRiello}.
The prototypical such example is the total electric charge contained in a region  -- i.e. precisely the source of $\varphi$.

%====================
\subsection{Horizontal symplectic potential with matter}\label{sec:matter}

Without further ado, let us introduce matter fields.
To start with, both $\Phi_c$ and $\theta$ are accordingly modified: e.g. in YM with charged Dirac fermions  $$\psi \in \mathcal C^\infty(\Sigma, W\otimes \bb C^4)$$ ($W$ is the fundamental representation of $G$ and $\bb C$ are the complex numbers), configuration space is now given by $\Phi_c=\{(A_i,\psi)\}$ and\footnote{Due to the presence of matter, that we consider to be of unit charge for simplicity, we have reintroduced a conventional $4\pi$ factor. We have also chosen conventions that make $\theta$ real-valued.}
\be
\theta = \tfrac{1}{4\pi } \theta_\text{YM} + \theta_\text{Dirac},
\qquad
\text{where}
\qquad
\theta^\text{Dirac} = -\tfrac{i}{2} \int \sqrt{g} \,\Big(\bar \psi \gamma^0 \dd \psi - \text{c.c.}\Big)
\ee 
 ($\gamma^\mu$ are Dirac's $\gamma$-matrices).
Without altering our definition of $\varpi=\varpi_{\perp}(\dd A)$, it is straightforward to check that $\theta$ can be decomposed in analogy to \eqref{eq:decomposition}
\begin{align}
\theta &= \tfrac{1}{4\pi } \theta^\perp_\text{YM} + \theta^\perp_\text{Dirac} + \tfrac{1}{4\pi } \int \sqrt{g} \,\tr\Big(\D^i\varphi_\perp \D_i \varpi_{\perp} + 4\pi \rho\varpi_{\perp} \Big), \notag\\
& \approx \tfrac{1}{4\pi } \theta^\perp_\text{YM} + \theta^\perp_\text{Dirac} + \tfrac{1}{4\pi } \oint \sqrt{h}\, \tr\Big(\D_s \varphi\, \varpi_{\perp} \Big)
\end{align}
where we used $\dd_H \psi := \dd \psi + \varpi \psi $, and introduced the Lie-algebra valued (color-)charge density 
\be
\rho = \tfrac{i}{2} \Big(\bar\psi \gamma^0 \tau^\alpha \psi - \text{c.c.}\Big) \tau_\alpha,
\ee
as well as the Gauss constraint with sources
\be
4 \pi\rho \approx \D_i E^i = - \D^2 \varphi_\perp.
\ee 
From these formulas one sees that all the properties derived in the absence of matter still hold even in its presence.
In particular, the difference between the SdW-horizontal and standard symplectic potentials is given precisely by the Coulombic piece of the electric field and the pure gauge part of $\dd A$, i.e. $\varpi_\perp$.

Again, although $\varphi^\#$ is clearly vertical, it does not carry the interpretation of a gauge component of $E$ ($E$ is gauge invariant in electrodynamics). Instead, another sort of decomposition ensues: the instantaneous states  of the charges completely characterize the instantaneous states of the $\varphi_\perp$ component of $E$,  and have no bearing on the other component, $\dot A^\perp$ (which is still susceptible to the currents, according to Ampere's law). Thus, the characterization of one component as ``radiative'' and the other as ``Coulombic'' stands as in the previous subsection.

In electrodynamics, whereas SdW-horizontal perturbations of the photon field correspond to transverse photons, SdW-horizontal perturbations of the electron field  can be seen as a finite-region generalization of Dirac's dressed electron \cite{GomesRiello2018}. We will come back to this analogy -- and its deficiencies -- in section \ref{sec:dressing}.

%====================
\subsection{Horizontal charges}

We are now in the position to discuss the horizontal symplectic charges in the presence of matter.
The first observation is that generically, these charges also vanish, for the simple reason that the horizontal differentials are designed to annihilate vertical perturbations: 
\be
Q^H[\xi] := \bb i_{\xi^\#} \theta^H \equiv 0.
\ee 
However, this conclusion can be avoided in particular situations.

\begin{figure}[t]
		\begin{center}
			\includegraphics[scale=0.17]{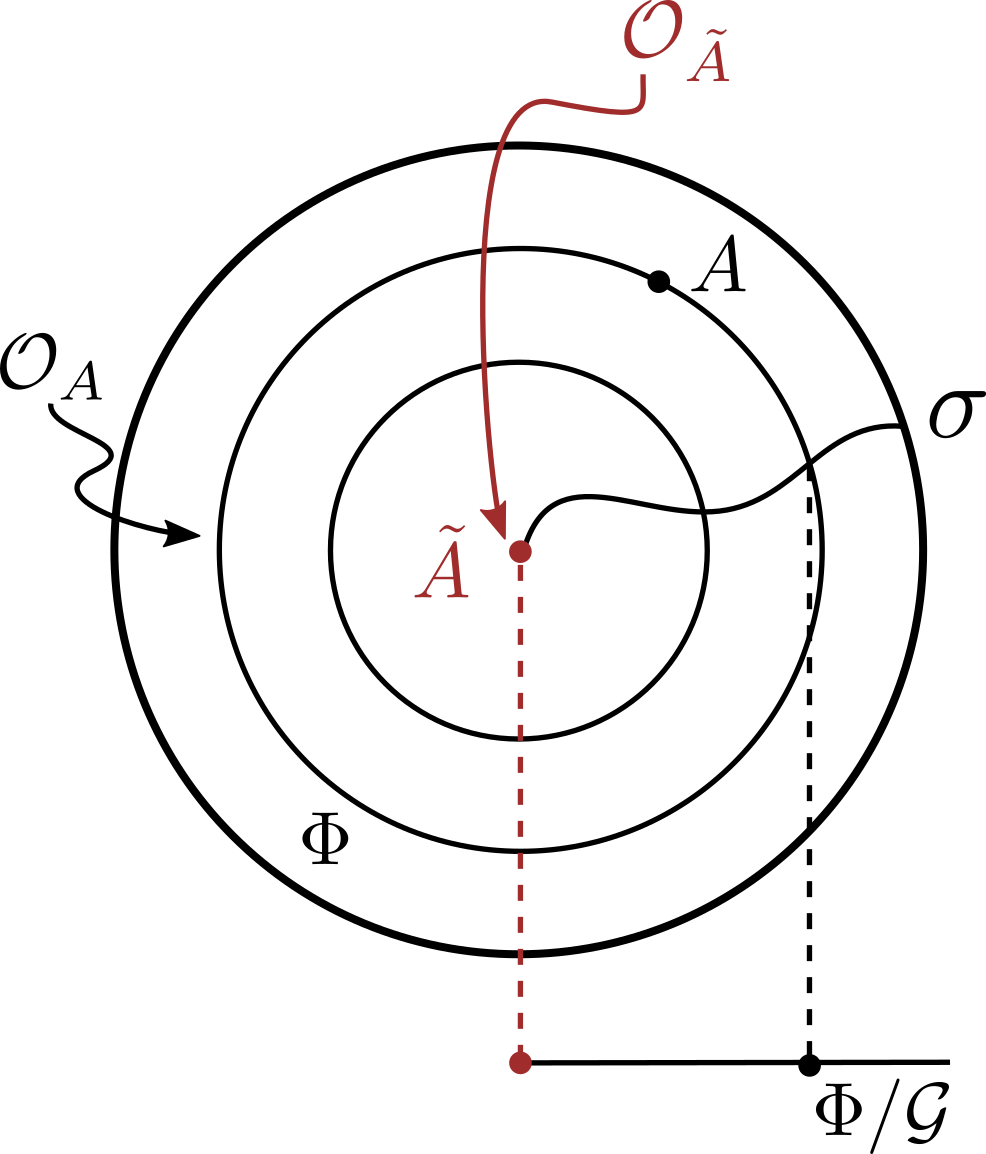}	
			\caption{In this representation $\Phi$ is the page's plane and the orbits are given by concentric circles; $\sigma$ is a section. The field $A$ is generic, and has a generic orbit, $\mathcal{O}_A$.  The field $\tilde A$ has a nontrivial stabilizer group (i.e. it has non-trivial reducibility parameters), and its orbit $\mathcal{O}_{\tilde A}$ is of a different dimension than $\mathcal{O}_A$. The projection of $\tilde A$ on  $\F/\G$ therefore sits at a qualitatively different point than that of $A$ (a lower-dimensional stratum of $\F/\G$).}
			\label{fig8}
		\end{center}
	\end{figure}

Consider a configuration $ \tilde A_i$ such that there exists a gauge transformation $\chi$ with the following property: 
\be
\delta_\chi \tilde A_i = \tilde \D_i \chi = 0.
\ee
Then, $\tilde A_i$ is said ``reducible'' and $\chi$ is called a ``reducibility parameter''. 
The reducibility parameters $\chi$ depend on the global properties of $\tilde A_i$ and  constitute a finite dimensional vector space (possibly of dimension zero). See figure \ref{fig8} for a representation of such fixed points  of an arbitrary field under a certain group action. 

 For non-Abelian theories reducible configurations are non-generic, in the same way as those spacetime metrics which admit Killing vector fields are non-generic.\footnote{Because of the existence of these configuration $\Phi_c$ is not quite a bona fide fibre bundle, and its base manifold is in fact a stratified manifold (see \cite{GomesHopfRiello} for a more thorough discussion of the geometry involved  and further references on the topic).}
In this respect, electromagnetism is an exception, since all its configurations have as a reducibility parameter the constant gauge transformation, $\chi^\text{EM}=const$.

Now, since $ (\bb i_{\chi^\#} \dd A_i)_{|\tilde A} = \delta_\chi \tilde A_i =  0 $, it immediately follows that at these configurations the projection property of $\varpi_{\perp}$ fails,\footnote{ More generally, at reducible configurations, the differential operator appearing in equation \eqref{eq:varpi_def} which defines $\varpi_{\perp}$ has a kernel, given by the reducibility parameters \cite{RielloSoft}. } since
\be
(\varpi_{\perp}(\chi^\#) )_{|\tilde A}= 0. 
\ee
However, in the presence of matter, this happens without $\chi^\#$ being strictly zero, since
\begin{align}
\xi^\#& = \int \d x \, \Big( (\D_i \xi)^\alpha(x) \frac{\delta}{\delta A_i^\alpha(x)} + (-\xi\psi)^B(x)\frac{\delta}{\delta \psi^B(x)}  \Big)\\
& \Rightarrow \chi^\# _{|\tilde A} =  \int \d x \, (-\chi\psi)^B(x)\frac{\delta}{\delta \psi^B(x)}  .
\end{align}
 where here $B$  is a spinorial index, e.g. $\gamma^\mu = (\gamma^\mu)^{B'}{}_B$.

Therefore, at reducible configurations, we have a finite number of  (non-trivially acting) gauge transformations that are also horizontal with respect to the SdW connection form. 
As such, we want to interpret this transformations as being ``physical'' and thus corresponding to physical -- rather than gauge -- symmetries.

This interpretation is further justified by the horizontal charges associated to $Q^\perp[\chi]$, which are easily verified to correspond to
\be
Q^\perp[\chi] := \bb i_{\chi^\#} \theta^\perp = \int_\Sigma \sqrt{g} \, \tr(\chi \rho)
\qquad\text{(at   $A = \tilde A$ and for $\chi$ such that $\tilde \D\chi =0$)}.
\ee
 In summary:
\be
Q^\perp[\xi] = \begin{cases}
0 & \text{if $A$ and $\xi$ are generic}\\
\int_\Sigma \sqrt{g}\, \tr( \chi \rho ) & \text{if $A=\tilde A$ is reducible and $\xi = \chi $, with $\tilde \D \chi = 0$ }
\end{cases}
\ee

In electromagnetism, where all configurations are reducible, the (only) horizontal Noether charge in $R$ is the total electric charge in $R$ -- and this does not need to vanish. 

Finally, notice that $\chi$ corresponds to a global symmetry of the background gauge field and there, if this symmetry happens to be preserved by the time evolution (recall that we are working in configuration space), the corresponding global charges will be conserved.  
This is analogous to the conserved quantities associated to Killing vector fields in general relativity. 

To conclude  this part of the paper, let us mention that a generalization of these results to (seemingly) more general asymptotic symmetries and charges, i.e. Strominger's leading soft charges \cite{strominger2018lectures}, has been proposed in \cite{RielloSoft}.

%========================================
\section{The composition of regions}\label{sec:gluing}

So far, we have  dealt with a single  functional connection $\varpi$, on both manifolds with and without boundaries. But to analyze the gluing of regions,  we must consider such boundaries to be \textit{fiducial}: not actual boundaries of the manifold.\footnote{Even if they are determined by particular properties of the fields themselves, the boundaries are not codimension one surface where ``space ends''---the standard interpretation of bounded manifolds.} This requires us to employ distributions, representing the idealized separating surfaces between such regions. 

Consider the following  setting: assume again that we take  Yang-Mills configuration space  to be $\F_{c}$: compactly supported  gauge fields over the simply connected manifold $\Sigma$, assumed here to be  \textit{without boundaries}. This entails that the relevant gauge transformations are also compactly supported. We embed in $\Sigma$  a separating surface (of codimension one), $S\subset\Sigma$, e.g. the hypersurface formed by the vanishing of one arbitrary (i.e. not necessarily  Cartesian) coordinate, $y=0$. 
For example, $S\simeq \bb R^{n-1}$, splitting $\Sigma\simeq \bb R^n$ into $\Sigma^+, \Sigma^-$ (e.g. regions of positive and negative $y$  coordinates), sharing a single boundary, $S$. 
We denote the respective configuration spaces, and all operators on them by $\pm$, as in $\F_c^\pm, \bb X^\pm,$ etc.
 (Notice that in previous sections we would have referred to either one of the $\Sigma^\pm$ as $R$.)

 Of course, more general situations can be envisaged. The simplest of which is the gluing of a ``shell'' around a finite region, implementing ``radial evolution''. This case can be treated with essentially the same techniques as here. The other obvious case is the gluing of finite regions across portions of their boundaries in such a way that ``corners'' are involved. This situation, on the other hand, would require a more careful study of  corner points. We leave the latter problem to future work, and focus on the simplest case of $\Sigma$ split in two regions $\Sigma^\pm$, joining at $\pp \Sigma^\pm \cong  S$ (up to orientations).

To formally encode the separation of regions, we introduce the Heaviside function in the half-maximum convention, on the $y$ direction, 
\be\label{eq:Heaviside}
\Theta_+(x^1,\cdots, x^{n-1},y):=\begin{cases}
0 \qquad \text{for}\qquad y< 0 \\
\frac12 \qquad \text{for}\qquad y=0 \\
1 \qquad \text{for}\qquad y> 0
\end{cases}
\ee
and similarly for $\Theta_-$. We abbreviate both by $\Theta_\pm(y)$.
 Accordingly, we have that\footnote{These formulas are also easily derivable by integrating against a compactly supported test function (as is standard with the space of tempered distributions), e.g.
 $$
\int \d y  \, \xi(y) \frac{d}{d y} \Theta_+(y)  :=  
- \int \d y \, \Theta_+(y) \frac{d}{d y} \xi(y) =  
- \int_{0^+}^\infty \d y  \, \frac{d}{d y} \xi(y) = \xi(0).
$$} 
\be
\frac{d}{dx^k}\Theta_\pm(y)=0\qquad \text{and}\qquad \frac{d}{dy}\Theta_\pm(y)=\pm \delta(y).
\ee

More generally, $\Theta_\pm$ are the characteristic functions of the regions $\Sigma^\pm$, with value $1/2$ at the boundary. Denoting $s_i$ the outgoing co-normal at $S$ with respect to the region $\Sigma^+$, one has\footnote{In local coordinates $(x^k,y)$ such that $S=\{ y=0\}$, $\delta_S(x^k,y) = \delta(y)$.}  $\partial_i \Theta_\pm= \pm s_i \delta_S$ which should be interpreted as a distributional expression encoded in the following equation for a smooth test vector field on $\Sigma$, $v^i(x)$:
\be
\int_{\Sigma^\pm} v^i \partial_i\Theta_\pm= \pm \oint_S s_i v^i .
\label{eq:dTheta}
\ee

Since the Heaviside function is constant in  field space, it applies equally to both  field-space points and their tangent vectors, so that\footnote{From here on we will often commit a slight abuse of notation, to simplify formulas; we conflate  field-space vectors with their components in  the standard coordinate system, namely: $\bb X=\int \bb X_i(x)\frac{\delta}{\delta A_i(x)}$. \label{ftnt:notation}}
\be\label{eq:heavi_dec} A=A_+\Theta_++A_-\Theta_-, \qquad\text{and}\qquad \bb X=\bb X_+\Theta_++\bb X_-\Theta_-.
\ee
If we suppose that $A^\pm\in \F_c^\pm$ join smoothly at $S$, as do $\bb X^\pm\in T_{A^\pm}\F_c^\pm$ since they all descend from a global smooth $A\in \F_c$ and $\bb X\in  T_{A}\F_c$, it means that we require continuity 
\be
\bb X_+(x,0)=\bb X_-(x,0).
\ee
This continuity condition suffices to get rid of the delta function which might emerge from $C^1$ smoothness:
\be \frac{d}{dy}\bb X=\frac{d}{dy}\bb X_+\Theta_++\frac{d}{dy}\bb X_-\Theta_-+(\bb X_+-\bb X_-)\delta(y).
\label{eq6}
\ee
Similarly, higher order continuity, and the absence of distributions at $S$, are ensured by the coincidence of derivatives of $\bb X_\pm$ at $S$.

%----------------------------------
\subsection{Regional horizontal gluing}

 Before proceeding, let us attend to a technical point.
 Embedding $\Sigma_\pm$ into $\Sigma$,   let $U_\pm$ be  open sets, $\Sigma_\pm \subset U_\pm\subset \Sigma$, which extend over an arbitrarily small neighborhood of $\Sigma_\pm$ within $\Sigma$.  We then define an extension of $\varpi_\pm$ as any  functional 1-form over $\Sigma$, $\tilde\varpi_\pm\in \Omega^1(\F_c, \fG)$,  which, for  $x\in \Sigma_\pm$, and for all smooth  field-space vectors  $\bb X_\pm\in  T_A\F_{U_\pm}\subset T_A\F_c$, with support over $U_\pm$ satisfies:  $\D^2 \tilde\varpi_\pm(\bb X)(x)=\D^i \bb X^\pm_i(x)$, and  $s^i\D_i\tilde\varpi_\pm(\bb X)=s^i\bb X_i$  at $S$ (and not $\pp U_\pm$). 

 In sum, if $\varpi_\pm := (\varpi_\perp)_\pm$ is defined as in \eqref{eq_SdW} for the bounded manifolds $\Sigma_\pm$, we are then constructing  arbitrary and smooth extensions of $\varpi_\pm$ to $U_\pm\supset \Sigma_\pm$.\footnote{In other words we consider arbitrary extensions of $\varpi_\pm$ to $U_\pm$, i.e. $\varpi_\pm=\int_{U^\pm} \d z \Delta^{-1}(z, x)\D^i \bb X_i(z)$, in the sense that their Green's functions reduce to the usual SdW connections on the bounded manifold for when $x,z\in\Sigma^\pm$.}
 These extensions are needed for mere technical reasons, and thus for what follows we just use $\varpi_\pm$ without the tildes.

Now, let us ask what are the conditions for gluing,  i.e. for reconstructing global horizontal data, from regional horizontal data.\footnote{ We assume throughout that the non-perturbative regional data is already smooth; i.e the base field $A$ is smooth, and decomposes as in \eqref{eq:heavi_dec}. }
To phrase this question mathematically, let us consider two smooth regionally horizontal perturbations $\bb h^\pm_i$ supported on $U^\pm$ respectively,  defined by the horizontal projection of the generic perturbation $\bb X^\pm$:
\be
\bb h^\pm := \bb X^\pm - \varpi_\pm^\# (\bb X^\pm)
\ee
and which must therefore satisfy, thanks to the defining properties of the SdW connection, the conditions analogous to \eqref{eq:sdw_A}
\be
\begin{cases}
\D^i\bb h^\pm_i=0 & \text{in } \Sigma^\pm\\
s^i\bb h^\pm_i = 0 & \text{at }S .	
\end{cases}
\label{eq:hor_conds}
\ee
See \cite{GomesHopfRiello} for a complete account. Moreover, we demand that these horizontals satisfy the following conditions at $S$:
\be
\D^{\phantom{+}}_{[i} \bb h_{j]}^+ = \D^{\phantom{+}}_{[i} \bb h_{j]}^- + \tfrac12[F_{ij}, \xi] \quad \text{at } S,
\label{eq:Dh_cont}
\ee
 since this ensures that the boundary is only fiducial, since it implies the matching of the perturbed field-strengths up to an infinitesimal\footnote{ Recall that, by hypothesis, the backgrounds configurations are supposed to glue smoothly across $S$.} gauge transformation. We will come back to it later.

 Finally, we introduce a global horizontal perturbation $\bb H$,  defined in the same manner as $\bb h^\pm$, but with respect to the global SdW connection $\varpi$ on the whole $\Sigma$.

\begin{figure}[t]
		\begin{center}
			\includegraphics[width=6cm]{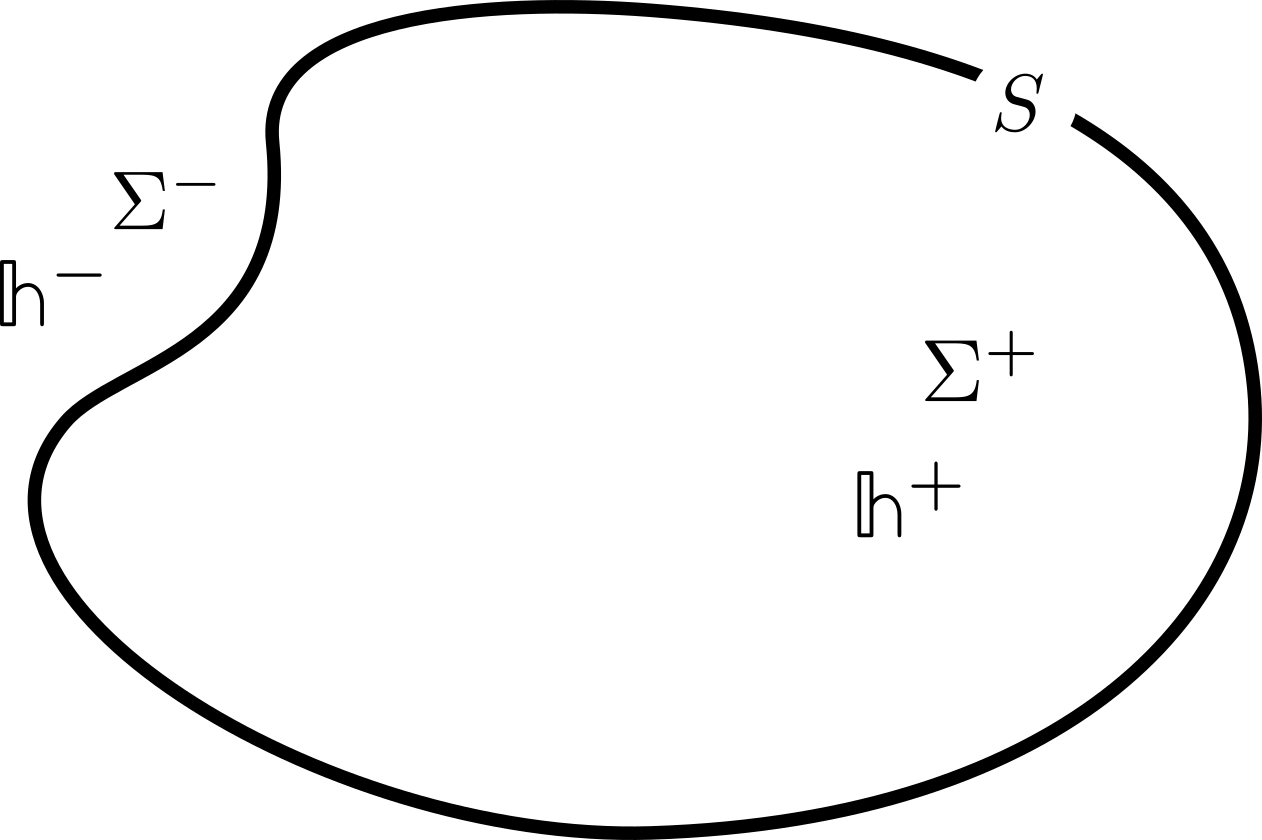}	
			\caption{ The two subregions of $\Sigma$, i.e. $\Sigma^\pm$, with the respective horizontal perturbations $\bb h^\pm$ on each side, along with the separating surface $S$. 
			}
			\label{fig4}
		\end{center}
	\end{figure}

The question we will now answer is: how do we construct from the regional $\bb h^\pm$ a global horizontal field $\bb H$ in a unique manner? In figure \ref{fig4}, we are given the two regional horizontal perturbations, and would like to uniquely construct a global, smooth horizontal perturbation. 

We will reconstruct the global horizontal field vector by adjoining to $\bb h^\pm$ an appropriate gauge transformation, functionally dependent solely on $\bb h^\pm$. Here it is very important to notice that {\it horizontal projections differ from gauge-fixings}: had we gauge-fixed, gauge-transformations  would have no longer been allowed to act non-trivially on either region. It is only by keeping gauge transformations around that we can compatibilize regional horizontal fields with a global horizontal field (see also appendix \ref{sec:dressing}). 

Hence, we want to define two smooth functions $\xi^\pm$ such that, given $\bb h^\pm$  satisfying \eqref{eq:hor_conds} and \eqref{eq:Dh_cont} the reconstructed field:
\be
\bb H=(\bb h^++\D\xi^+)\Theta_++(\bb h^-+\D\xi^-)\Theta_-
\label{eq:reconstructed_H}
\ee 
is smooth and globally horizontal (here we use simplified notation, see footnote \ref{ftnt:notation}).

 Let us chart a roadmap, consisting of three steps, on how we will go about finding such $\xi^\pm$:
 \begin{enumerate}
 
\item First, we will impose restrictions on the difference $\bb h_+-\bb h_-$ at the boundary. This restriction  is necessary for $\bb H$ to be smooth at the boundary.\footnote{ We will later (in section \ref{sec:int_h}) find that the restriction is equivalent to demanding `no intrinsic boundary horizontal modes', and can itself be relaxed with interesting consequences.} Combined with the horizontality of the regional $\bb h_\pm$, the requirement of smoothness gives us conditions on the longitudinal and transverse derivatives of $(\xi^+-\xi^-)$ at the boundary. One condition states the equality of $s^i\D_i \xi_\pm{}_{|S}$, while  the longitudinal condition allows us to solve for the boundary tangential difference $(\xi^+-\xi^-)_{|S}$ in terms of $(\bb h_+ - \bb h_-)_{|S}$. 
\item  Second, since now $\bb H$ is smooth, we impose its horizontality and obtain another condition on the bulk of $\xi_\pm$'s, It states that, in their own regional domains, they should be covariantly harmonic.
\item As the last step, we employ the previous facts for the unique reconstruction of $\xi_\pm$ in terms of $\bb h_\pm$. To do so, we massage the above equations  into (\textit{i}) two homogeneous elliptic boundary-value problems, one per region, with generalized Neumann conditions (as encountered in the study of $\varpi_\perp$), together with (\textit{ii}) an equation fixing the generalized-Neumann boundary condition in terms of $(\xi^+-\xi^-)_{|S}$. The latter equation involves the inversion of an elliptic operator intrinsic to the boundary and the action of Dirichlet-to-Neumann operators associated to each region.
\end{enumerate}

For the first step, we observe that the continuity condition \eqref{eq:Dh_cont} is satisfied if and only if the curvature perturbation of $\bb H$ is smooth across $S$. 
As before, in \eqref{eq:heavi_dec}, first order continuity  demands that 
\be
 (\bb h^+ - \bb h^-)_{|S} =\D(\xi^+-\xi^-)_{|S},
\label{eq:continuity}
\ee
which in turn merely identifies $\xi$ in \eqref{eq:Dh_cont} with $(\xi^+-\xi^-)_{|S}$. 
 Now, we can decompose this equation into its transverse and longitudinal components with respect to $S$. Since the component of $\bb h^\pm$ transverse to $S$ vanishes because of regional horizontality \eqref{eq:hor_conds}, we conclude that 
\be
s^i\D_i(\xi^+ - \xi^-)_{|S} = 0,
\label{eq:normalmatchingxi}
\ee
i.e. that the normals derivatives of $\xi^\pm$ at $S$ must match; and that $(\xi^+-\xi^-)_{|S}$ is solely determined by  two horizontals'  mismatch  at the boundary:
 \be
 (\xi^+-\xi^-)_{|S} = - (\D^2_S)^{-1}\D^c_S(  \bb h_c^+ - \bb h_c^- )_{|S},
 \label{eq:Shoriz}
 \ee
 where $\D_S$ is the  gauge- and space-covariant differential intrinsic to $S$ and $\D^2_S$ its Laplacian,  and we have denoted the index contraction intrinsic to $S$ with an early latin letter, $c$ (not to be confused with the indices used for the Lie algebra basis).
 The functional space on which this intrinsic differential acts is $\Phi^S_c$, i.e. the space of the pullbacks onto $S$ of the fields in $\Phi_c$.  This concludes the first step of the proof, outlined above.
 
Now we move to the second step: assuming that the whole $\Sigma$ has no boundaries, with $\bb H$ given by \eqref{eq:reconstructed_H}, by smearing the condition $\D^{i} \bb H_{i}=0$ signifying  global horizontality:  
 \be 
 \int_\Sigma \,\tr\Big[\eta\, \D^{i}\left((\bb h_{i}^+ + \D_{i}\xi^+)\Theta_++(\bb h_{i}^-+\D_{i}\xi^-)\Theta_-\right)\Big]=0
 \ee
 for any $\eta\in C^\infty(\Sigma,\Lie(G))$.
 Now, since $\D^{i}\bb h_{i}^\pm=0=s^{i} \bb h_{i}^\pm$, and thanks to the identity $\pp_{i}\Theta_\pm=\pm s_{i}\delta_S$ \eqref{eq:dTheta}, we get:
  \be 
  \int_{\Sigma^+} \tr\Big[ \eta\,\D^2\xi^+ \Big] + \int_{\Sigma^-}\tr\Big[\eta\,\D^2\xi^-\Big] + \oint_S \tr\Big[\eta\, s^{i}\D_{i}\left(\xi^+-\xi^- \right)\Big]=0,
 \ee
 where the vanishing of last term above was  already deduced  by \eqref{eq:normalmatchingxi}.  
 Since $\eta$ is arbitrary, we can obtain a last, condition, on the bulk, described in step 2 of the outline.
We conclude that the $\xi^\pm$  satisfy the following elliptic boundary value problem
\be\label{lambda}
\begin{cases}
\D^2 \xi^\pm = 0 &\text{in }\Sigma^\pm\\ 
s^i\D_i\left(\xi^+-\xi^- \right)=0 &\text{on }S\\
(\xi^+-\xi^-)_{|S} = -  (\D^2_S)^{-1}\D^c_S(  \bb h^+_c - \bb h_c^- )_{|S} & \text{on }S
\end{cases}
\ee
This concludes the second step. Now we must use the appropriate PDE tools  to show that this boundary value problem determines $\xi_\pm$ (essentially) uniquely in terms of the  regional horizontal fields $\bb h^\pm$.

For the third step,  we start by setting
\be
\Pi := s^i(\D_i \xi^\pm){}_{|S},
\label{eq:Pidef}
\ee
from the second equation in \eqref{lambda}.  
Once we obtain a closed form for  $\Pi$, we can determine $\xi^\pm$  by solving the boundary value problem given by \eqref{eq:Pidef} and the first equation of \eqref{lambda}. 
In this way, given $\Pi$, the $\xi^\pm$ will be uniquely determined up to elements of the stabilizers of $A_{|\Sigma^\pm}$ respectively. In the topologically simple case  that we are analyzing so far, this will be the only ambiguity present in the reconstruction -- more on this later in section \ref{sec:gluing_matter}. 
 To determine $\Pi$, we then introduce generalized \textit{Dirichlet-to-Neumann operators} (see \cite{BFK} and references therein), $\mathcal{R}_\pm$, mapping -- in each region -- Dirichlet conditions for a (covariantly) harmonic function to the corresponding (gauge-covariant) Neumann conditions.  But let us be more explicit.
 
{
 The Dirichlet-to-Neumann operator $\mathcal R$ is defined by
 \be
 \label{eq:DTN}
 \mathcal R(u):=s^i\D_i (\zeta_{u})_{|S},
 \ee
for a manifold with boundary $S$  and outgoing normal $s^i$, and for the unique  (covariantly) harmonic Lie-algebra-valued function $\zeta_{u}$  defined by the elliptic Dirichlet boundary value problem: $\D^2 \zeta_{u}=0$ with $(\zeta_{u})_{|S } = u$. We note that the subscript $u$ here encodes the Dirichlet boundary condition we have employed. If we use  superscripts to denote (gauge-covariant) Neumann boundary conditions, then by definition we would have $\zeta^{\mathcal{R}(u)} \equiv \zeta_u$. In fact, since the corresponding Neumann problems  also have unique solutions,\footnote{Again: due to the possible presence of stabilizers, this is not strictly true, but this fact does not alter our conclusions on $\bb H$. See below.} $\cal R$ is invertible, i.e. $\zeta^\Pi \equiv \zeta_{\mathcal R^{-1}(\Pi)}$.

We thus define $\cal R_\pm$  associated to $\Sigma^\pm$ with boundaries $\pm S$, and their inverses $\mathcal{ R}_\pm^{-1}$.
 
 From \eqref{eq:Pidef} and the covariant harmonicity of  $\xi$ is itself (from the first equation of \eqref{lambda}), we have 
\be\label{eq63}
 \xi^\pm= \zeta_{(\pm)}^{ \pm\Pi}=\zeta^{(\pm)}_{\mathcal{R}_\pm^{-1}(\pm \Pi)}.
\ee
Here, the sub(super)script $(\pm)$, indicates the domain of the functions, i.e.  that the respective functions are defined over $\Sigma_\pm$ respectively. 
We will now proceed by fixing  $\Pi$ uniquely through the last equation of \eqref{lambda}.

  By the properties of $\mathcal{R}$ we have
\be  
\mathcal{R}^{-1}_\pm({ \pm}\Pi) = {\pm} \mathcal{R}^{-1}_\pm(\Pi).
\label{eq:signimportant}
\ee 
The  $\pm$ sign inside the argument of $\mathcal{R}^{-1}$  is due to the fact that, at $S$, $s^i\D_i \xi^+ = s^i\D_i\xi^- $ but $s^i$ is the outgoing normal on one side, and the ingoing normal on the other, so they fix opposite Neumann conditions on the two sides. Therefore, together with \eqref{eq63} and \eqref{eq:signimportant} we obtain
\be
(\xi^+-\xi^-)_{|S}=\mathcal{R}^{-1}_+(\Pi)- \mathcal{R}^{-1}_-(-\Pi)=\Big(\mathcal{R}^{-1}_+  + \mathcal{R}^{-1}_-\Big)(\Pi).
\ee
 That is: we obtain a relation between the (gauge-covariant) Neumann boundary $\Pi$ and the difference of the Dirichlet boundary conditions $\xi_\pm{}_{|S}$. 
 
  Finally, with this relation in hand, we can  provide a formula that fixes $\Pi$, in terms of  $(\bb h^+ - \bb h^-)_{|S}$, thus we conclude the third and last step of our proof.
To get to this final conclusion, we replace the formula above into last equation of \eqref{lambda}, obtaining: 
 \be
\Big(\mathcal{R}^{-1}_+  + \mathcal{R}^{-1}_-\Big)(\Pi) =  (\D^2_S)^{-1}\D^c_S(  \bb h_c^+ - \bb h_c^- )_{|S}.
\label{eq:RRPi}
\ee
which is the equation  $\Pi$ must satisfy. Its solution fixes $\xi^\pm$ uniquely through \eqref{eq63}, and summarizes the solution to the entire set of equations \eqref{lambda}.}

As a consisteny check of our result, we show that the operator $(\mathcal{R}^{-1}_+  + \mathcal{R}^{-1}_-)$ is invertible (and therefore the solution is unique). Invertibility  follows from the $\cal R_\pm$  being a positive self-adjoint operator, and from the relative sign appearing on the left-hand-side of \eqref{eq:RRPi}.

Now we therefore shift our attention to a proof   that the generalized Dirichlet-to-Neumann operators $\cal R_\pm$ are self-adjoint and have positive spectrum. Again $\zeta_u\neq0$ denotes the unique solution to the problem $\D^2 \zeta_u = 0$ in the bulk with boundary condition $(\zeta_u)_{|S}=u$. Then, given any Lie-algebra valued functions $u, v$ on the boundary,
\begin{align}\label{eq:self_adj}
0\leq \int_{\Sigma^+} g^{ij}\tr( \D_i \zeta_u \D_j \zeta_v ) 
 & = - \int_{\Sigma^+} \tr( \zeta_u \D^2 \zeta_v )+ \oint_{\partial \Sigma} s^i \tr( \zeta_u \D_i\zeta_v )\notag\\ 
 & = \oint_{S}  \tr (u \mathcal R_+(v))
 = \oint_{S} \tr (\mathcal  R_+(u) v) 
\end{align}
(where we have omitted the volume densities). The first step in \eqref{eq:self_adj} consists in an integration by parts and applying the following identity, valid for any smearing $\eta \in C^\infty(\Sigma,\Lie(G))$:
\be
\tr\Big( - \eta\,  \pp^i\D_i \zeta  + g^{ij}[A_i,\eta] \D_j \zeta \Big) = 
\tr\Big( - \eta\, \pp^i\D_i \zeta  - g^{ij} \eta [A_i, \D_j \zeta] \Big) =
\tr\Big( - \eta\, \D^2 \zeta  \Big) .
\ee
Setting  $u=v$ in \eqref{eq:self_adj} we obtain positivity:
$$ \oint_{S} \tr (u \mathcal R_+(u))>0
$$
 Similar manipulations lead to the analogous conclusion for $\mathcal R_-$.
 
  To summarize, there is a unique choice of $\xi^\pm$ available for the gluing of horizontals $\bb h_\pm$. They are found by solving \eqref{lambda}, giving a covariant harmonic function with a specific Neumann boundary condition: 
\begin{align}  \xi^\pm = \zeta_{(\pm)}^{ \pm\Pi}\quad\text{with}\quad \Pi=\Big(\mathcal{R}^{-1}_+  + \mathcal{R}^{-1}_-\Big)^{-1}\left(  (\D^2_S)^{-1}\D^c_S(  \bb h_c^+ - \bb h_c^- )_{|S}\right)
\end{align}
This concludes the argument. 
 
\subsubsection*{Ambiguities}
 As we mentioned, the above proof also shows that the only possible kernels correspond to the reducibility parameters  $ \chi^\pm$ of  regionally reducible configurations $ \tilde A^\pm := A{}_{|\Sigma^\pm}$ ($ \chi^\pm$, $ \tilde D^\pm \chi^\pm =0$, is what we identified  in the previous sections as ``symmetries''). 
Let us recall that the existence of these reducibility parameters -- and there are at most $\mathrm{dim}(\Lie(G))$ of them -- depends on the configuration $A$ one is studying.
They give rise to a finite dimensional kernel for $\mathcal R$.
The ensuing ambiguity arising when trying to invert $\mathcal R$ can  -- in the absence of matter -- be fixed arbitrarily, since it does not affect the reconstruction of $\bb H$, which depends on $\D \xi$, and not $\xi$ itself.

But we should note there are two possible exceptions to the uniqueness of the gluing: (\textit{i}) the presence of stabilizers of the gauge field in conjunction to the presence of matter, and (\textit{ii}) a non-trivial  topology of $\Sigma$. In fact, when the topology is nontrivial,  the room for  discrepancy is equivalent to broken cohomology cycles. 
 We will  discuss these points after having further examined in section \ref{sec:int_h} the reconstruction \eqref{eq:Shoriz} of $(\xi^+ - \xi^-)_{|S}$ subsumed by equations \eqref{eq:Dh_cont} and \eqref{eq:continuity}.

\subsection{Boundary horizontal modes}\label{sec:int_h}
Two remarks are in order. 
First of all, as the reconstruction is essentially unique,\footnote{ Even at reducible configurations in the presence of matter, the ambiguity is at most finite dimensional and therefore it also cannot correspond to the presence of new boundary degrees of freedom.} we conclude that there is no need to add new information, nor degrees of freedom,  to be able to glue regional and horizontal perturbations (at least for trivial topology,  but see  section \ref{sec:topology}).
The boundary gauge parameter $(\xi^+-\xi^-)_{|S}$ can be understood as partially parametrizing the ignorance of what lies on the other side.

Second, there is a more constructive procedure to deal with the continuity condition \eqref{eq:Dh_cont} -- the condition which implements the fiducial status of the boundaries. 
This procedure, that we now discuss, can be iteratively applied to the ``boundaries of the boundaries'' and so on, opening a door to the discussion of the more general gluing schemes involving corners that we referred to at the beginning of this section. 

In a gauge theory, the space of the pullbacks to $S$ of the fields in $\Phi_c$ defines a new boundary  field space, $\Phi^S_c$ which is isomorphic to the space of gauge fields intrinsic to $S$. Moreover, the induced metric on $S$ defines a supermetric $\bb G^S$ on $\Phi^S_c$. From this, one can define an SdW connection $\varpi_{S,\perp}$ on $\Phi^S_c$.
Now, thanks to the second of the equations \eqref{eq:hor_conds},  $s^i\bb h_i=0$, the difference between two {\it generic}\footnote{I.e. that do {\it not} have to satisfy the continuity condition \eqref{eq:Dh_cont}.} horizontal perturbations {$\bb h^\pm$} defines,  without any loss of information, a vector field intrinsic to the boundary:
\be\label{eq:bbx}
\bb x := \iota^*_S(\bb h^+ - \bb h^-) \in \mathrm T_{\iota_S^* A} \Phi^S_c.
\ee
This vector field can be decomposed via $\varpi_{S,\perp}$ into its horizontal and vertical parts \textit{within} $\Phi^S_c$:
\be\label{dec_S} \bb x=\bb h_S+\D_S\xi
\ee
Given equations \eqref{eq:bbx} and \eqref{dec_S}, then it becomes clear that the continuity condition for a fiducial boundary, \eqref{eq:continuity}, is equivalent to demanding that $\bb x$ has no horizontal component, i.e. $\bb h_S=0$ (in which case, of course, the $\xi$'s of  \eqref{dec_S} and \eqref{eq:Dh_cont} are identified).

From these simple observations, we conclude that $S$ can be considered a fiducial boundary if and  only if $\bb x$ is purely boundary-vertical, that is {\it a smooth global and horizontal $\bb H$ exists only if $\bb x = \varpi_{S,\perp}^\#(\bb x)$.} In such a case, this last equation is only a more formal way to write \eqref{eq:continuity}, with $(\xi^+-\xi^-)_{|S}=-\varpi_{S,\perp}(\bb x)$ being a rewriting of  \eqref{eq:Shoriz}.

 Summarizing, we showed that, under the assumption that the $\bb h^\pm$ satisfy conditions \eqref{eq:hor_conds} and \eqref{eq:Dh_cont} -- in which case there are no horizontal modes intrinsic to the boundary itself -- and for non-reducible configurations, we can construct a unique global horizontal field from gluing, namely $\bb H$. This global field depends solely on $\bb h^\pm$, and yet generically does not restrict to $\bb h^\pm$ in each region (unless $\xi^\pm\equiv0$). 

This says that whenever regional horizontal perturbations are even in principle ``gluable", they will fully determine the manner in which they can be glued. If they cannot be glued, then there is no corresponding smooth global horizontal perturbation that treats the boundary as fiducial.
In this case, gauge invariant quantities, such as the electromagnetic field, have to be non-continuous at the boundary. In electromagnetism this corresponds to a non-vanishing boundary density of charge or current: in other words, the boundary has to be physically determined by a physical  perturbation, that can be unambiguously identified. 

 In conjunction with the gluing results, the intrinsic nature of the difference between regional horizontals, as expressed in \eqref{eq:bbx}, bears an interesting possibility. To see it,  note that if $S$ itself had corners, i.e. if it was subdivided into regions $S^\pm$ sharing a boundary, we could have repeated the same treatment for two possible horizontal differences, $\bb h^+_S-\bb h_S^-$, themselves arising from the difference of horizontals one level up, as expressed in \eqref{dec_S}.  I.e. given boundaries of boundaries, etc, termed $\Sigma_n$ (e.g. $\Sigma=\Sigma_0, S=\Sigma_{1}$), up to the exact terms $\D_{\Sigma_{n+1}}\xi$ used for gluing, we can repeat the treatment we have here developed for the descending horizontal differences, $\bb h_{\Sigma_n}^+-\bb h_{\Sigma_n}^-$. This procedure suggests an accompanying chain of descent for $\bb h$ and $\d$, which we plan to explore in the future.

%==================
\subsection{Gluing with matter\label{sec:gluing_matter}}

In the presence of matter, gluing is more subtle. 
Let us first introduce some notation. Let $\bb h^\pm = \bb h^\pm_A + \bb h^\pm_\psi$ and $\bb H = \bb H_A + \bb H_\psi$, where
\be \begin{cases}
\bb H_A = (\bb h_A^+ + \D \xi^+) \Theta_+ + (\bb h_A^- + \D \xi^-) \Theta_-
\\~\\
\bb H_\psi = (\bb h_\psi^+ - \xi^+\psi) \Theta_+ + (\bb h_\psi^- +  \xi^-\psi) \Theta_-
\end{cases}.
\ee
Here, we are implicitly using the SdW connection to assess horizontality. It is important to note that the matter horizontal components $\bb h^\pm_\psi$ are then, in a sense, parasitic on the gauge-field: they are just the corrected velocity of the matter sector, by the vertical displacement of the gauge sector, namely,  $\bb h_\psi=\bb X_\psi-\varpi(\bb X_A)\bb X_\psi$; they need not satisfy equations of their own. 

Then,  $\bb H$ (and $\bb h^\pm$) is horizontal (regionally horizontal, respectively) if and only if $\bb H_A$ ($\bb h^\pm_A$) is.
This means in particular that the above procedure aimed at the determination of $\xi^\pm$ is completely insensitive to the presence of matter, and can be applied in the same way.
The hypothesis of continuity of the original global perturbation $\bb X = \bb X_A+ \bb X_\psi$ ensures that the same $\xi^\pm$ needed to glue one field will work for the other  as well.

Now, this all works seamlessly, unless either one of the {\it regional} configuration of the gauge potential, i.e. $A_\pm$, admits a stabilizer: then, the reconstruction of the corresponding $\xi^\pm$ will be ambiguous in a way that will generically\footnote{This is always the case in QED, where we can always add constants $c^\pm$ to the reconstructed $\xi^\pm$ and where a constant phase shift will affect the Dirac fermions, unless they vanish. In a non-Abelian theory, the zoology is more complicated, and will depend on the gauge group as well as the type of matter fields (fundamental, adjoint, etc).} affect the matter fields in a nontrivial way  (recall that these ambiguities have no effect on the reconstruction of the global horizontal gauge potential).

Let us suppose, for definiteness, that only the regional configuration $\tilde A^+ = A{}_{|\Sigma^+}$ is reducible  by a single reducibility parameter $\chi^+$, while $A^- = A{}_{|\Sigma^-}$ is not. Generalizations are straightforward.
Then, we face two distinct possibilities: either $\chi^+$ stabilizes $\psi^+:=\psi {}_{|\Sigma^+}$ {\it at} the boundary -- that is $\chi^+ \psi^+{}_{|S} = 0$ -- or it does not. 

In the first case, which can materialize e.g. if $\psi^+$ vanishes there, gluing of the two perturbations $\bb h^\pm$ is still possible but will give rise to {\it physically distinct global perturbations}.  To see this, we explicitly parametrize the 1-parameter family of solutions for $\xi^+$, by choosing an origin $\xi^+_o$ and writing the general solution in the following way (where $r$ depends on the charge group)
\be
\xi^+_r := \xi^+_o + r \chi^+
\qquad r \in \bb R \text{ or } \bb C.
\ee
 Then, the corresponding global horizontal perturbations  $\bb H^r = \bb H_A^r + \bb H_\psi^r$ are given by
\be
\bb H_A^r \equiv \bb H^o_A
\qquad\text{and}\qquad
\bb H_\psi^r = \bb H^o_\psi  - r \chi^+ \psi \,\Theta_+.
\ee
Each of the $\bb H^r$  for different $r$ is horizontal -- hence physical, according to our identification -- but distinct  (unless the even more restrictive condition is given that $\chi^+$ stabilizes $\psi^+$ throughout $\Sigma^+$).
This setup formalizes 't Hooft's beam splitter thought experiment \cite{Brading2004}, and can be used to provide a concrete example for the considerations of Wallace and Greaves,  characterizing ``symmetries with direct empirical significance'' \cite{GreavesWallace}.
We plan to investigate this setup in more detail in the future.

The second case, on the other hand, allows us to glue the two perturbations together if and only if we can find an $r$ such that
\be
\xi^+_r \psi{}_{|S} = \xi^- \psi{}_{|S}.
\ee
With the continuity hypothesis for the original global field perturbation $\bb X = \bb X_A + \bb X_\psi$, this equation would then fix the global ambiguity. 

 These compatibility requirements between $\chi^+$ and $\psi^+$ could be further formalized in terms of the kernel of the Higgs functional connection introduced in \cite{GomesHopfRiello}. We refer the reader to this reference for further details and considerations.

In the following, we will put aside the more subtle case of gluing reducible configurations in the presence of matter, and focus on the generic case only.

%=================

\subsection{Example:  1 dimensional gluing and the emergence of topological modes}\label{sec:topology}

In this section we work out a simple example, implementing the gluing of 1-dimensional intervals. 
Two cases are given: two closed intervals are glued into a larger closed interval, and one closed interval is glued on itself to form a circle. 
This second case falls outside the topologically trivial setup we adopted for the rest of the paper. Nonetheless,  this case allows us to easily discuss, without introducing a host of new technologies, the emergence of new global (or ``topological'') degrees of freedom associated to the non trivial cohomology of the circle.

\subsubsection*{Gluing into  an interval}
Let us start by  considering two closed intervals $I^+=[0,1]$ and $I^-=[-1,0]$, that we shall glue together to form  a new closed interval $I = [-1,1]$.
 We shall see that, since on the interval the gauge potential must be pure gauge, the regional horizontal perturbations must vanish -- a fact consistently encoded by our gluing formula. Although somewhat trivial, this example helps us set the stage for the gluing into a circle.
 
We first characterize the 1-dimensional gauge fields and their horizontal perturbations.
One dimensional gauge fields are always locally pure gauge,
\be
A^\pm = g_\pm^{-1} \d g_\pm
\ee 
for $g_+(x) = \mathrm{Pexp}\int_0^x A$ on $I^+$ and similarly on $I^-$, where we choose $g_-$ such that $g_-(0)=\bb 1$ too ($x=0$ is where the gluing takes place).
Since in one dimension $s\cdot \bb h{}_{|S}=0$ implies $\bb h_{|S}=0$, SdW-horizontal perturbations $\bb h^\pm$ in $I^\pm$, according to  \eqref{eq:hor_conds} must satisfy the equations 
\be
\D^\pm \bb h^\pm =0
\qquad\text{and}\qquad
\bb h^\pm{}_{|\pp I^\pm} = 0,
\ee which can be rewritten in terms of $\tilde{\bb h}{}^\pm := g_\pm \bb h^\pm g^{-1}_\pm$  as $\pp\tilde{\bb h}{}^\pm = 0$ and $\tilde{\bb h}{}^\pm{}_{|\pp I^\pm} = 0$. 
Now, these equations can be solved to give $\tilde{\bb h}{}^\pm = 0$ and hence
\be
\bb h^\pm=0.
\label{eq:1dhoriz}
\ee
This is solely an immediate consequence of the pure gauge character of all 1-dimensional configurations, and therefore all perturbations over topologically trivial regions must be purely vertical.

At this point we can move on and analyze the gluing. Again, the global horizontal vector is denoted by 
\be
\bb H=(\bb h^++\D\xi^+)\Theta_++(\bb h^-+\D\xi^-)\Theta_-
\ee
as in \eqref{eq:reconstructed_H}. The relevant equations for gluing arise as in  \eqref{lambda}, with a couple of new features:  (\textit{i}) there is no analogue of the last equation of \eqref{lambda}; $\bb h_i$ has only one component that is transverse to the zero-dimensional gluing surface $S$,  and  (\textit{ii}) since the total horizontal vector has two endpoint boundaries, we have to add one equation per global boundary of the interval $I=[-1,1]$.  

Thus, here the smoothness and horizontality of $\bb H$ imply the following conditions on $\xi^\pm$:
\be
\label{eq:gluing1d}
\begin{cases}
\D^2 \xi^\pm = 0 &\text{in }I^\pm\\ 
\D \left(\xi^+-\xi^- \right)=0 &\text{at } \pp I^+ \cap \pp I^-=\{0\}\\
\D \xi^\pm=0 &\text{at } \pp I = \{\pm 1\}
\end{cases}
\ee
Now, again, by defining $\tilde \xi{}^\pm := g_\pm \xi^\pm g_\pm^{-1}$, we can turn the covariant derivatives into ordinary ones. This allows us to readily solve these equations. In fact, the bulk equations (the first of \eqref{eq:gluing1d}) tell us that 
\be
\tilde \xi^\pm = \pm \tilde \Pi^\pm x + \tilde\chi^\pm,
\ee
where $\tilde\chi^\pm$ are constant functions valued in $\Lie(G)$ corresponding to two arbitrary reducibility parameters of the vanishing configuration $\tilde A^\pm=0$. This is a concrete example of the discussion in the previous section.

 Now, the second equation of \eqref{eq:gluing1d} sets $\tilde\Pi^+ = - \tilde\Pi^-$, and the third one sets them equal to zero. Since $\tilde \chi_\pm$ don't affect the value of the regional horizontal fields, we hence conclude that in this case the unique solution to the gluing problem at hand is $\xi^\pm = 0$ which readily leads to $\bb H =0$, consistently with the general regional result \eqref{eq:1dhoriz}. This concludes the gluing of two closed intervals, $I^\pm$, into a larger one, $I=[-1,1]$.

\subsubsection*{Gluing into circle}
We now move on to the second case, where one interval, $I=[-\pi,\pi]\ni\phi$, has its ends  glued to form a unit circle. To keep the two cases notationally distinct, we have denoted an element of the circle by $\phi$, as opposed to $x$ of the interval in the previous case.
This case requires a little more care. 

The idea is to split $I$ into two intervals which overlap around $\phi=0$, e.g. on the interval $U_\epsilon:=(-\epsilon,\epsilon)$. I.e. let $I^- = \left[ - \pi , \epsilon\right)$ and $I^+ = \left(-\epsilon, \pi\right]$, so that we can glue at $\phi=\pm\pi$ according to the procedures of the above section, while matching the overlap of charts  around $\phi=0$ to close the interval into a circle. 

This allows us to separate the problem of gluing from the problem of covering the circle. The latter is accomplished  by overlapping open charts,  with transition functions which appropriately match the gauge configuration.

Let us start by analyzing the background configuration $A^\pm$ on $I^\pm$. We assume, as in the previous sections, that the configurations $A^\pm$ join smoothly at $\phi=\pm\pi$.\footnote{It would be more appropriate to introduce new coordinates to glue at a specific value of the coordinate. We persist  in this slightly sloppier, but more compact, language, merely flagging the possibility of confusion. }

As above, $A^\pm$ are pure gauge, i.e. $A^\pm = g^{-1}_\pm \d g_\pm$ with $g_+(\pi)=g_-(-\pi)$. 
On the other hand, on $U_\epsilon$, the configurations $A^\pm$ do not have to  be equal; they need only be related by the action of a gauge transformation $f$, the transition function.  Since we are in 1-dimension, this does not constitute a restriction; one simply has $f= g_-^{-1} g_+$.

Now, we move on to consider the horizontal perturbations.
We shall find that the relevant horizontality equations for $\bb h^\pm$ involve boundary conditions only at $\phi = \pm \pi$, and the one for $\bb H$ does not involve boundary conditions at all. In particular no boundary conditions are imposed at the open-extrema of the intervals $I^\pm$. This is not  because the intervals are open, but rather because those boundaries do not exist for the global $\bb H$ : they are just artificial boundaries of the charts used to describe the circle. For this reason, the SdW inner product through which horizontality is defined does not see them. But let us show this more constructively.

We start from the observation that on the overlap region $U_\epsilon$, generic perturbations $\bb X^\pm$  must be gauge related through $\bb X^+ = \Ad_f \bb X^-$.
 This means that there is no difficulty nor ambiguity in the patching (using the appropriate partitions of unity over $U_\epsilon$) of the SdW inner products over $I^+$ and $I^-$ to obtain an inner product over $I$  between two perturbations $\bb X^\pm$ and $\bb Y^\pm$ satisfying the overlap condition we have just described. Recalling that SdW-horizontality is the requirement of being orthogonal with respect to the SdW supermetric to any purely vertical vector, we hence see that the horizontality condition for $\bb H$ does {\it not} involve boundary conditions at the non-glued boundaries of $I^\pm$, i.e. at $\phi=\pm \epsilon$. Of course, this was an expected result from the closed nature of the manifold on which $\bb H$ resides.

 Focusing now on horizontal perturbations, it is easy to see that this discussion doesn't change the fact that $\bb h^\pm=0$, since the manifold on which they reside still  has boundaries  at $\phi=\pm \pi$. Note moreover that  $\bb h^\pm=0$ implies that their matching on $U_\epsilon$ is  automatic. However, this discussion leads us to a horizontality condition for $\bb H$ that is distinct from the one found for the gluing into an interval \eqref{eq:gluing1d}. Indeed, in the present case, we find 
\be
\label{eq:gluing1d-circle}
\begin{cases}
\D^2 \xi^\pm = 0 &\text{in }I^\pm\\ 
\D \left(\xi^+-\xi^- \right)=0 &\text{at } \phi=\pm\pi\\
\end{cases}
\ee
with {\it no} extra conditions at $\phi=\pm\epsilon$.
Hence, it is readily clear that the solution for $\xi^\pm$ is here much less restrictive than it was in the closed interval case considered above: in this case we find that  %{\old\bf CHECK ACTION}
\be
\xi^\pm =  {g_\pm^{-1}} (\tilde \Pi \phi + \tilde \chi^\pm) {g_\pm}.
\ee 
 with the same, possibly non-vanishing, $\tilde \Pi$, and therefore %{\old\bf CHECK ACTION}
\be
\bb H = {g_\pm^{-1}} \tilde \Pi {g_\pm}.
\ee
As for the background, matching the perturbed configurations in $U_\epsilon$ comes at no cost (since $\bb h_\pm =0$).

In summary, we see that the gluing procedure has no unique solution in this case, as a consequence of the  absence of a second ``outer'' boundary for the interval (which is glued into a circle). The second outer boundary is instead replaced by the chart matching.\footnote{ The decoupling of chart transitioning and horizontal gluing can be made into a more general feature. For instance, had we wished to cut up the circle into three segments, we would divide the interval $[0,2\pi]$ into three sets, $I_1=[0,2\pi/3], I_2=[2\pi/3, 4\pi/3], I_3=[4\pi/3, 2\pi]$, with $\bb h_i\in I_i$. Then we can cover the circle with three charts $U_{1,2,3}$,  given in larger, but largely overlapping, domains: $D_1=[0,4\pi/3], D_2=[\pi/3, 2\pi], D_3=[4\pi/3, \pi/3]$. Then $\bb h_1$ and $\bb h_2$ glue entirely within the $U_1$ chart domain $D_1$; $\bb h_2$ and $\bb h_3$ similarly glue in $D_2$; and $\bb h_3, \bb h_1$ glue in $D_3$.  In this way, one decouples the chart matching from the horizontal gluing; we can cyclically glue all $\bb h_i$'s first and find the appropriate chart transition later, independently. In that case, it is the cyclicity of the equations that yields one less condition. This type of concatenating construction can be extended to higher dimensional manifolds.}
 We thus obtain a  one-parameter family of solutions parametrized by an element $\tilde \Pi\in\Lie(G)$.
This element constitutes the perturbation of the Wilson-loop observable around the circle (Aharonov-Bohm phase), which is precisely the unique physical degree of freedom present there. 
The existence of this new topological mode is of course related to the non-trivial homological properties of the circle. 
 Also note that in our formalism this topological mode arises automatically and originates solely from the interplay between gauge and topology.

As a closing remark we point out that in the absence of charged matter, this topological mode must be conjugated to the electric field present on the interval and detectable at the boundary of $I$. The detailed relationship between the two on topologically non-trivial domains requires an extension of the results of section \ref{sec:thetaH} (and \ref{sec:Gauss} in particular) and is left for future work.

\section{Conclusions}\label{sec:conclusions}

 There were  main themes to previous work  \cite{GomesRiello2016, GomesRiello2018, GomesHopfRiello, GomesStudies, RielloSoft}  on the  functional connection form. Namely: (\textit{i}) that it  provides a regional decomposition of field perturbations into physical and gauge, (\textit{ii}) that it  identifies gauge-invariant, regional charges without the need for gauge-fixing, (\textit{iii}) that it  provides a straightforward notion of gluing,  and (\textit{iv}) that it provides a general and unifying notion of dressing (see appendix \ref{sec:dressing}).

After reviewing the basic technology of the framework in section \ref{sec:field_space}, this paper  advanced two main innovations: (\textit{1})  as most of the previous work thus far was pursued in the Euclidean covariant context, we here expanded the results of \cite{GomesHopfRiello} to the $D$+1 decomposition,  with the goal of  properly accounting for time evolution; (\textit{2}) whereas in previous work gluing was treated at an abstract level, here we derived the explicit formulae dictating its behaviour. The novelty of the paper therefore lies  in sections  \ref{sec:thetaH} and \ref{sec:gluing}, treating points 1 and 2 above, respectively.

\subsubsection*{ The SdW-connection and symplectic geometry in configuration space}

Regarding the first innovation, in treating the $D$+1 properties of the framework, we found that the choice of the Singer-DeWitt connection, argued to be naturally inherited from the Lagrangian of the theories in question, has another useful property: the associated symplectic form is the only one to split the purely radiative modes  from the instantaneous ones determined by the Gauss law. In other words, the special property of the SdW $\varpi_\perp$ we found in the present paper is the separation between Coulombic and radiative contributions to the symplectic form.    This newly found property strengthens our argument for themes (\textit{i}) and (\textit{ii}), above.

 Indeed, the horizontal-vertical decomposition as implemented by the SdW connection matches the Helmholtz decomposition of vector fields in the Abelian case, but it provides a natural  field-dependent generalization thereof in the non-Abelian case, and it reveals this standard decomposition as just one  incarnation of a richer geometrical structure on  configuration space.
 Moreover, even in the Abelian case, we are unaware of previous literature describing the consistent gluing of the Helmholtz decomposition -- a project we undertook in section \ref{sec:gluing}  -- and  the relationship of the decomposition to Noether charges -- which we covered in section \ref{sec:thetaH}.

However, as the saying goes, there is no free lunch: our constructions and results come at a price. In the Euclidean  picture, everything is manifestly covariant, and the price to pay is the  arbitrariness in the choice of a functional connection where none\footnote{Even if a covariant analogue of the SdW connection introduced here could be argued more natural also in that context.} is canonically given. In configuration space, on the other hand, the SdW connection we discussed in this article is canonically selected by its compatibility relation with the symplectic potential -- a compatibility rooted in the fact that both of them spur from the kinetic supermetric on configuration space. The price to pay, in this case, is the manifest breaking of Lorentz symmetry, i.e. the canonical SdW connection we construct is anchored on a specific choice of foliation. 
This tie between gauge and Lorentz symmetry is reminiscent of \cite{FrohlichMorchioStrocchi1979, BalachandranVaidya2013}.

 In describing the gained properties of the horizontal symplectic potential and generators, we also stressed the importance of considering field-dependent transformations. Why should gauge transformations be field-independent? If one is committed to the view that gauge is solely descriptive redundancy, we see no reason. In fact, in the absence of boundaries, nothing is gained or lost by this extension. On the other hand, incorporating  field-dependent transformations serves as a great diagnostic tool for many of the problems besetting the formulation of symplectic charges and generators in bounded regions. It also further explains the efficacy of the  functional connection-form in dealing with all of these issues. 
 
  It is worthwhile to note a disparity between the emergence of the boundary terms here and in \cite{DonnellyFreidel}. Indulging in a slight abuse of nomenclature, here our boundary terms arise as a by-product of the complete split of ``Coulombic'' and ``radiative'' modes. The radiative part is completely gauge-invariant, but $\theta$ is as it was: still gauge-\textit{variant}. In \cite{DonnellyFreidel}, no such split is performed; an extra contribution to the boundary degrees of freedom is posited to render a modified symplectic potential gauge invariant.   
  
  In our case, there is a parallel with  reduction processes in gauge theories,  which proceed in two steps: (\textit{a}) solve the constraints, (\textit{b}) reduce. These steps correspond to solving the constraint and then removing $\varphi$ from $\theta$ in \eqref{eq:symp_pot}. This parallel can be gleaned as underlying section \ref{sec:Gauss}.
 
The following list summarizes the meaning and gluing properties of quasilocal physical degrees of freedom we found in this paper. More specifically, it summarizes the standing of relations between the SdW choice of connection and boundary/bulk information necessary for gluing, with a focus on the role of the Coulombic component $\varphi_\perp$:
\begin{enumerate}
\item Off-shell of the Gauss constraint, $\varphi_\perp$ is an independent variable, unrelated to the matter content or  the boundary-local electric flux $E_S = s^i E_i$. It is only isolated by the geometric horizontality/verticality construction. If  $\varphi$ is dropped from the description at this level, i.e. off-shell of the Gauss constraint, information  is lost, both locally and globally.

\item  On-shell of  the Gauss constraint: $\varphi_\perp$ is determined by the  regional matter content and $E_s$. The latter ingredient summarizes the relevant nonlocal information coming from the outside of the region of interest. If there is no outside, there is no $E_s$. Similarly, if there is no outside, the $\varphi_\perp$ piece to the presymplectic potential automatically drops out.\footnote{In both bounded and closed cases, $\theta_\perp$ is gauge invariant and solely expressed in terms of horizontal quantities (this morally corresponds to $\theta_\perp$ being symplectic rather than presymplectic).} Therefore, $\theta_V=\theta-\theta_\perp$ can be seen as encoding the  (presymplectic) pairing of the information from the outside of the region, i.e. $E_s$, with the pure gauge piece of $\delta A$, i.e. $\varpi(\delta A)$.  These statements are not automatic; they are achieved by using the SdW connection.  They also support the idea that the significance of gauge lies in the coupling of subsystems \cite{RovelliGauge2013}.

\item  On-shell of the Gauss constraint, in both the regional and the global cases, $\varphi_\perp$ is just “re-evaluated” at every step and in every region,   in terms of the regional $E_s$ and matter content.  In this sense, gluing regional information requires only the knowledge of the regional horizontal modes. 

\end{enumerate}

\subsubsection*{ The SdW-connection and gluing}

 Moving to the second part of this work,  in section \ref{sec:gluing}, we  extended the work on gluing of \cite{GomesHopfRiello}. We showed here for the first time that --  in topologically-trivial domains -- regional, or quasi-local horizontal  perturbations contain the necessary and sufficient information for gluing into a global, smooth, and purely horizontal mode, if and only if  the perturbations do not introduce any currents intrinsic to the boundary, i.e. if and only if the boundary is  and remains fiducial.  This is in contrast to the wide-spread belief that extra data is always necessary at the boundary \cite{DonnellyFreidel, Geiller:2017xad, Speranza:2017gxd}.
 
Gluing requires an adjustment of each regional horizontal perturbation, so that the joined field is smooth and itself horizontal. The adjustment is along the gauge direction in each region, and is only licensed because we are neither gauge-fixing nor quotienting the gauge degrees of freedom from the theory, therefore each regional horizontal field can still be acted on by gauge transformations.  We further discuss the relation between gauge-fixing and the connection-form in appendix \ref{sec:dressing}.

 In topologically trivial domains, the appropriate regional gauge transformations used for gluing are unique; they are given by a closed exact formula exploiting the properties of the Dirichlet-to-Neumann operator (but again, we impose no particular boundary conditions on the underlying fields). Moreover, the vertical adjustments are explicit functionals of the mismatch between the original regional horizontals at the common boundary.  
  The role of horizontal modes intrinsic to the boundary was considered in section \ref{sec:int_h}, but much remains to be done. In particular, it seems that a chain of descent could apply for boundaries of boundaries, etc. We expect it to be useful when discussing more complex patterns of gluings, especially into global domains with non-trivial cohomology.

Importantly, non-trivial cohomological cycles give rise to  ambiguities in the gluing procedure, which we review below. Another type of ambiguity arises  from the presence of stabilizers of the fields. This latter ambiguity is irrelevant in the absence of matter, and is independent of the topology; we studied it in section \ref{sec:gluing_matter}. It gives rise to the possibility of symmetries with ``direct empirical significance'' \cite{GreavesWallace} which we plan to study in an upcoming paper.

In the topologically non-trivial case, we found that cohomology cycles can be encoded in the underdetermination of the global horizontal modes from the regional ones. In a simple 1-dimensional example, we  showed that precisely the global Aharonov-Bohm  mode arises in this space. 
 This provides the concrete mechanism within our framework through which regional pure-gauge modes can surge to physical status in the presence of a non-trivial cohomology. Such modes would of course always be finite in number.
We conjecture that the entire difference between regional and global modes arises thus; from non-trivial cohomology, and can also be recovered from our gluing procedure. In the general case, we expect that a central role in this reconstruction procedure will be played by the descent chain of SdW connections discussed above.

Finally, we observe that the operator $(\mathcal R_+^{-1} + \mathcal{R}^{-1}_-)$ coincides, in the Abelian case, with the operator featuring in the BFK formula for the composition of (zeta-regularized) functional determinants of Laplacians \cite{BFK, KirstenBFK}. Following the considerations of \cite{Agarwal}, this suggests that the present construction should be relevant for the calculation of the entanglement entropy in gauge theories -- {where fiducial boundaries are crucial, but not easily implementable in previous set-ups (see e.g. the ``brick wall'' of \cite{Donnelly:2014fua, AronWill}).}

We leave a complete study of the descent chain, of the gluing into topologically nontrivial domains, of entanglement entropy, and of a generalization of all these to general relativity and diffeomorphism symmetry (see \cite{GomesRiello2016,Alinew}) to future work.

%==========================================

\subsection*{Ackowledgements}
 We are thankful to Florian Hopfm\"uller for detailed comments on the first version of this draft, and to Ali Seraj for  valuable feedback.
Research at Perimeter Institute is supported in part by the Government of Canada through the Department of Innovation, Science and Economic Development Canada and by the Province of Ontario through the Ministry of Economic Development, Job Creation and Trade.  HG was supported by the Cambridge International Trust.

\appendix
%==========================================
\section{Time dependent gauge transformations\label{app:A0gauge}}

To properly understand time dependent gauge transformations and the role of the term $\varpi(\bb v)$ appearing in $A_0$, we need to clarify the geometrical meaning of the quantities  appearing in the relevant equations.

First of all,  $\varphi$ has to be included in $\Phi$, giving rise to $\Phi' = \Phi\otimes\{\varphi\}$. The extent to which $\varphi$ is fully constrained and drops from the horizontal symplectic structure is explained in section \ref{sec:Gauss}.

Then, it is important to appreciate that the infinitesimal gauge transformations $\delta_\xi A_i$, $\delta_\xi \dot A_i$ and $\delta_\xi A_0$, have to be understood as field-space Lie derivatives of {\it  functionals}  on $\Phi$ along the vertical vector field $\xi^\#$.
This vertical vector field can be both field dependent, $\dd \xi \neq 0$, and time dependent, $\pp_t \xi \neq 0$. We reserve the partial derivative for the derivative of $\xi$ at some fixed point in $\Phi$.

Given an evolution flow $\frac{\d}{\d t}$ in configuration space, the function $\dot A_i$ is understood to be the {\it total} derivative of $A_i$ (understood as a coordinate function on $\Phi'$) along this flow:
\be
\dot A_i := \frac{\d}{\d t} A_i.
\ee
This function, constitutes the components of the  configuration-space ``velocity'' field 
\be
\bb v = \int \sqrt{g} \, \Big( \frac{\d}{\d t} A_i^\alpha\Big)(x) \frac{\delta}{\delta A_i^\alpha(x)}.
\ee
where (hereafter, $\bb v(\xi) := \bb i_{\bb v}\dd \xi$)
\be
\dot \xi := \pp_t \xi + \bb v(\xi)
\ee

With this, it is immediate to compute 
\be\label{eq:delta_xi_dotA}
\delta_\xi \dot A_i := \bb L_{\xi^\#}  \frac{\d}{\d t} A_i = \frac{\d }{\d t} (\D_i \xi) =  \pp_i \dot\xi  + [\dot A_i,\xi] + [A_i ,\dot \xi] = \D_i \dot \xi + [\dot A_i ,\xi].
\ee 
Notice that this the derivative along $\xi^\#$ of the function $\dot A_i$, and {\it not} the component of\footnote{Here, $\lbr\cdot,\cdot\rbr$ denotes the Lie bracket between vector fields in $\mathrm T\Phi$.} $\bb L_{\xi^\#} \bb v \equiv \lbr \xi^\#, \bb v\rbr $, which instead is equal to
\be
\lbr \xi^\#, \bb v\rbr = \int \sqrt{g}\, \Big( \xi^\#(\dot A_i^\alpha)  - \bb v(\D_i\xi)^\alpha\Big)(x) \frac{\delta}{\delta A^\alpha_i(x)} = (\dot \xi)^\# - \bb v(\xi)^\#,
\ee
 where we used \eqref{eq:delta_xi_dotA} for $\xi^\#(\dot A_i) = \delta_\xi \dot A_i$, as well as
\be
\bb v(\D_i\xi) = \bb v( \pp_i \xi + [A_i, \xi]) =  [ \dot A_i ,\xi] + \D_i \bb v(\xi).
\ee 

From the definition \eqref{eq:A0} of $A_0$ as 
\be
A_0 := \varphi + \varpi(\bb v) \qquad \text{with} \qquad \bb L_{\xi^\#} \varphi := [\varphi, \xi],
\ee
the defining properties \eqref{eq:varpi_def} of $\varpi$ -- i.e. $\varpi(\xi^\#) = \xi$ and $\bb L_{\xi^\#} \varpi = [\varpi, \xi] + \dd \xi$, -- as well as the above identities, it readily follows that
\begin{align}
\delta_\xi A_0 &= \bb L_{\xi^\#} \varphi + (\bb L_{\xi^\#} \varpi)(\bb v) + \varpi(\lbr \xi^\#, \bb v\rbr)\notag\\
& = [\varphi, \xi]+ [\varpi(\bb v), \xi] + \bb v( \xi) + \varpi( (\dot \xi)^\# - \bb v(\xi)^\#)\notag\\
& = [\varphi + \varpi(\bb v), \xi] + \dot \xi  = \D_0 \xi.
\label{eq:A0proof}
\end{align}

Finally, we prove that these transformations combine to give the expected transformation property to $E_i = \dot A_i - \D_i \varpi(\bb v)$:
\begin{align}
\delta_\xi E_i &= \frac{\d}{\d t} (\D_i \xi)  - \Big(  [\delta_\xi A_i, \varpi(\bb v)]  + \D_i (\bb L_{\xi^\#}\varpi)(\bb v) + \D_i \varpi(\lbr\xi^\#, \bb v\rbr) \Big) \notag\\
& = \Big( \D_i \dot \xi + [\dot A_i , \xi] \Big) +\notag\\ & \quad - \Big([\D_i \xi, \varpi(\bb v)]  + \D_i ( [\varpi(\bb v), \xi] + \bb v(\xi) ) + \D_i \varpi( (\dot \xi)^\# - \bb v(\xi)^\#) \Big) \notag\\
& =    [\dot A_i ,  \xi] -   [\D_i  \varpi(\bb v), \xi] ) =  [E_i, \xi]
\end{align}
where we used the projection property \eqref{eq:varpi_def} of $\varpi$, i.e. $\varpi(\eta^\#) = \eta$ for any Lie-algebra valued $\eta$.

 To avoid confusions, we re-iterate that $\delta_\xi E_i = \bb L_{\xi^\#} E_i$  is (very!) different from $\bb L_{\xi^\#} \bb E = \lbr \xi^\#, \bb E\rbr$. Indeed, a short computation shows that $ \lbr \xi^\#, \bb E\rbr =- \bb E(\xi)^\#$, where $\bb E(\xi)$ stands for the functional derivative of $\xi$ along $\bb E$. Both $\bb L_{\xi^\#} E_i$ and $\lbr \xi^\#, \bb E\rbr$ are Lie-derivative along $\xi^\#$, but the first is the Lie derivative of a function on $\Phi$, while the second is the Lie derivative of a vector field on $\mathrm T\Phi$.

We conclude this appendix by noticing that equation \eqref{eq:A0proof} shows the announced result that $A_0$ has automatically the correct transformation properties once we assume that $\varphi$ transforms in the adjoint representation: the burden of instilling $A_0$ with its typical non-homogenous transformation property under time- (and field-)dependent gauge transforamtions is fully carried by the term built out of $\varpi$.

%==========
 \section{Relation to gauge-fixing and dressings \label{sec:dressing}}

As mentioned in section \ref{sec:radiative}, in the Abelian case, given an initial configuration, we can ``extend'' the horizontal decomposition from the infinitesimal $\bb X\in T_A\F_c$ to the full field  configuration, $A_i$. 

Such an extension suggests a parallel between the SdW connection and Dirac's notion of dressing -- but crucial differences between the two exist and should not be neglected.
Roughly, Dirac's idea consists of introducing a gauge-invariant version of the electron field by ``dressing'' the bare electron field through a nonlocal cloud of photons in the form of a field-dependent ``gauge transformation'', 
\be
\hat \psi(x) := e^{ie \int \frac{\d y}{4\pi} \frac{\pp^i A_i(y)}{|x-y|} }\psi(x).
\ee
Being gauge invariant, this operator can be associated to a ``physical'' electron. The specific form of the dressing was chosen so that the associated operator acting on the vacuum creates not only the electron at $x$, but also its electrostatic field throughout space.
Now, the analogy with $\varpi_{\perp}$ is the following: taking $\Sigma$ to be $\bb R^3$ and ignoring boundary conditions, the dressing factor on top of the exponential is nothing but $\Delta^{-1} \pp^i A_i$, which means that \cite{GomesRiello2018}
\be
\delta \hat \psi = e^{i e \int \frac{\d y}{4\pi} \frac{(\pp^i A_i)(y)}{|x-y|} } \delta_{H,{\perp}} \psi
\qquad
\text{where}
\qquad
\delta_{H,{\perp}} \psi := \delta \psi +\varpi_{\perp} \psi,
\ee
Similarly, by dressing the photon field by the same field-dependent gauge transformation, one readily obtains the gauge-fixed transverse photon field $\hat A$ and an analogous relation between $\delta \hat A$ and $\delta_{H,{\perp}} A$.

This relation between horizontal perturbations and dressings is intriguing, but should not be taken too literally. 

First, using the  functional connection does not correspond to performing a gauge-fixing: vertical motions in  configuration space are still allowed and can actually be assessed -- and possibly factored out -- through the use of the connection form. In particular, through $\varpi$, it is possible to define a  configuration-space analogue of the Aharonov-Bohm phase, which would be instead completely lost in a gauge-fixed description. Treatments like the one of \cite{HollandsBMS} seem to suggest that this might be the correct language to geometrically capture the memory effects revived by Strominger and collaborators (see the discussion of ``historical dressings'' of \cite{GomesHopfRiello}). 

Such a distinction between a horizontal/vertical split and a gauge-fixing becomes vivid in the Hamiltonian case. There, a gauge-fixing requires the introduction of second class constraints. If conjugate variables are used to eliminate this set of second-class constraints, nothing is left to harmonize between different regions, and the work done in section \ref{sec:gluing} would not have been possible.

Second, this simple relation between dressings and $\varpi_{\perp}$ is lost in the non-Abelian case.
In fact, the proper generalization of the dressing factor to non-Abelian gauge theories involves the construction of Wilson-lines along paths in configuration space that link a reference field configuration, $A^\star_i$, to the configuration to be ``dressed'', $A_i$. This construction is {\it non-local in configuration space}.
In the Abelian case, the role of this field-space nonlocality is lost for two reasons: the flatness of the Abelian $\varpi_{\perp}$ makes the path dependence irrelevant, while the ``standard'' choice $A^\star_i=0$ hides the (crucial) dependence of ``dressings'' on a reference configuration.
Dressings as field-space Wilson-lines for $\varpi$ are related to other constructions present in the literature, e.g. the Gribov-Zwanziger framework and the Vilkovisky-DeWitt geometric effective action. We refer to \cite{GomesHopfRiello} for a more thorough analysis (see also \cite{francoisthesis}).

\bibliographystyle{bibstyle_aldo}
%\bibliography{references}

\begin{thebibliography}{10}
\providecommand{\url}[1]{\texttt{#1}}
\providecommand{\urlprefix}{URL }
\newcommand{\eprint}[1]{\hypersetup{urlcolor=BrickRed}\href{http://arxiv.org/abs/#1}{\tt arxiv:#1}\hypersetup{urlcolor=MidnightBlue}}




\bibitem{Polikarpov}
{P.~V. Buividovich} \protect\BIBand{} {M.~I. Polikarpov},
\newblock 2008
  \hypersetup{urlcolor=MidnightBlue}\href{http://dx.doi.org/10.1016/j.physletb.2008.10.032}{\textit{{Entanglement
  entropy in gauge theories and the holographic principle for electric
  strings}}}\hypersetup{urlcolor=MidnightBlue}.
\newblock Phys. Lett.
\newblock B670 141\hypersetup{urlcolor=BrickRed} [\eprint{0806.3376}]
  \hypersetup{urlcolor=MidnightBlue}

\bibitem{Casini_gauge}
{H. Casini}, {M. Huerta}, \protect\BIBand{} {J.~A. Rosabal},
\newblock 2014
  \hypersetup{urlcolor=MidnightBlue}\href{http://dx.doi.org/10.1103/PhysRevD.89.085012}{\textit{Remarks
  on entanglement entropy for gauge fields}}\hypersetup{urlcolor=MidnightBlue}.
\newblock Phys. Rev. D
\newblock 89 085012
\newblock [\eprint{1312.1183}]

\bibitem{GiddingsDonnelly}
{W. Donnelly} \protect\BIBand{} {S.~B. Giddings},
\newblock 2016
  \hypersetup{urlcolor=MidnightBlue}\href{http://dx.doi.org/10.1103/PhysRevD.94.104038}{\textit{{Observables,
  gravitational dressing, and obstructions to locality and
  subsystems}}}\hypersetup{urlcolor=MidnightBlue}.
\newblock Phys. Rev.
\newblock D94(10) 104038\hypersetup{urlcolor=BrickRed} [\eprint{1607.01025}]
  \hypersetup{urlcolor=MidnightBlue}

\bibitem{AbrahamMarsden}
{R. Abraham} \protect\BIBand{} {J.~E. Marsden}, 1985.
\newblock \textit{Foundations of Mechanics}.
\newblock
\newblock Addinson-Wesley

\bibitem{GomesRiello2016}
{H. Gomes} \protect\BIBand{} {A. Riello},
\newblock 2017
  \hypersetup{urlcolor=MidnightBlue}\href{http://dx.doi.org/10.1007/JHEP05(2017)017}{\textit{{The
  observer’s ghost: notes on a field space
  connection}}}\hypersetup{urlcolor=MidnightBlue}.
\newblock JHEP
\newblock 05 017\hypersetup{urlcolor=BrickRed} [\eprint{1608.08226}]
  \hypersetup{urlcolor=MidnightBlue}

\bibitem{GomesRiello2018}
{H. Gomes} \protect\BIBand{} {A. Riello},
\newblock 2018
  \hypersetup{urlcolor=MidnightBlue}\href{http://dx.doi.org/10.1103/PhysRevD.98.025013}{\textit{Unified
  geometric framework for boundary charges and particle
  dressings}}\hypersetup{urlcolor=MidnightBlue}.
\newblock Phys. Rev. D
\newblock 98 025013

\bibitem{GomesHopfRiello}
{H. Gomes}, {F. Hopfmüller}, \protect\BIBand{} {A. Riello},
\newblock 2019
  \hypersetup{urlcolor=MidnightBlue}\href{http://dx.doi.org/https://doi.org/10.1016/j.nuclphysb.2019.02.020}{\textit{A
  unified geometric framework for boundary charges and dressings: Non-Abelian
  theory and matter}}\hypersetup{urlcolor=MidnightBlue}.
\newblock Nuclear Physics B
\newblock 941 249
\newblock [\eprint{1808.02074}]

\bibitem{GomesStudies}
{H. Gomes},
\newblock 2019
  \hypersetup{urlcolor=MidnightBlue}\href{http://dx.doi.org/https://doi.org/10.1016/j.shpsb.2019.04.002}{\textit{Gauging
  the boundary in field-space}}\hypersetup{urlcolor=MidnightBlue}.
\newblock Studies in History and Philosophy of Science Part B: Studies in
  History and Philosophy of Modern Physics
  \newblock[\eprint{1902.09258}]

\bibitem{RielloSoft}
{A. Riello}, 2019.
\newblock \textit{{Soft charges from the geometry of field space}}
  \hypersetup{urlcolor=BrickRed} [\eprint{1904.07410}]
  \hypersetup{urlcolor=MidnightBlue}

\bibitem{Barnich}
{G. Barnich} \protect\BIBand{} {F. Brandt},
\newblock 2002
  \hypersetup{urlcolor=MidnightBlue}\href{http://dx.doi.org/10.1016/S0550-3213(02)00251-1}{\textit{{Covariant
  theory of asymptotic symmetries, conservation laws and central
  charges}}}\hypersetup{urlcolor=MidnightBlue}.
\newblock Nucl. Phys.
\newblock B633 3\hypersetup{urlcolor=BrickRed} [\eprint{hep-th/0111246}]
  \hypersetup{urlcolor=MidnightBlue}

\bibitem{DonnellyFreidel}
{W. Donnelly} \protect\BIBand{} {L. Freidel},
\newblock 2016
  \hypersetup{urlcolor=MidnightBlue}\href{http://dx.doi.org/10.1007/JHEP09(2016)102}{\textit{{Local
  subsystems in gauge theory and gravity}}}\hypersetup{urlcolor=MidnightBlue}.
\newblock JHEP
\newblock 09 102\hypersetup{urlcolor=BrickRed} [\eprint{1601.04744}]
  \hypersetup{urlcolor=MidnightBlue}

\bibitem{Geiller:2017xad}
{M. Geiller},
\newblock 2017
  \hypersetup{urlcolor=MidnightBlue}\href{http://dx.doi.org/10.1016/j.nuclphysb.2017.09.010}{\textit{{Edge
  modes and corner ambiguities in 3d Chern–Simons theory and
  gravity}}}\hypersetup{urlcolor=MidnightBlue}.
\newblock Nucl. Phys.
\newblock B924 312\hypersetup{urlcolor=BrickRed} [\eprint{1703.04748}]
  \hypersetup{urlcolor=MidnightBlue}

\bibitem{Speranza:2017gxd}
{A.~J. Speranza},
\newblock 2018
  \hypersetup{urlcolor=MidnightBlue}\href{http://dx.doi.org/10.1007/JHEP02(2018)021}{\textit{{Local
  phase space and edge modes for diffeomorphism-invariant
  theories}}}\hypersetup{urlcolor=MidnightBlue}.
\newblock JHEP
\newblock 02 021\hypersetup{urlcolor=BrickRed} [\eprint{1706.05061}]
  \hypersetup{urlcolor=MidnightBlue}

\bibitem{Camps}
{J. Camps},
\newblock 2019
  \hypersetup{urlcolor=MidnightBlue}\href{http://dx.doi.org/10.1007/JHEP01(2019)182}{\textit{{Superselection
  Sectors of Gravitational Subregions}}}\hypersetup{urlcolor=MidnightBlue}.
\newblock JHEP
\newblock 01 182\hypersetup{urlcolor=BrickRed} [\eprint{1810.01802}]
  \hypersetup{urlcolor=MidnightBlue}

\bibitem{HKT}
{S.~A. Hojman}, {K. Kuchar}, \protect\BIBand{} {C. Teitelboim},
\newblock 1976
  \hypersetup{urlcolor=MidnightBlue}\href{http://dx.doi.org/10.1016/0003-4916(76)90112-3}{\textit{{Geometrodynamics
  Regained}}}\hypersetup{urlcolor=MidnightBlue}.
\newblock Annals Phys.
\newblock 96 88

\bibitem{Giulini_gauge}
{D. Giulini},
\newblock 1995
  \hypersetup{urlcolor=MidnightBlue}\href{http://dx.doi.org/10.1142/S0217732395002210}{\textit{{Asymptotic
  symmetry groups of long ranged gauge
  configurations}}}\hypersetup{urlcolor=MidnightBlue}.
\newblock Mod. Phys. Lett. A
\newblock A10 2059\hypersetup{urlcolor=BrickRed} [\eprint{gr-qc/9410042}]
  \hypersetup{urlcolor=MidnightBlue}

\bibitem{kobayashivol1}
{S. Kobayashi} \protect\BIBand{} {K. Nomizu}, 1963.
\newblock \textit{Foundations of differential geometry. {V}ol {I}}.
\newblock
\newblock Interscience Publishers, a division of John Wiley \& Sons, New
  York-Lond on

\bibitem{Giulini_diff}
{D. Giulini}, 1995.
\newblock 
 \textit{Notes on Differential geometry}.
\newblock Unpublished
\newblock [\hypersetup{urlcolor=BrickRed}\href{https://qig.itp.uni-hannover.de/~giulini/papers/Diffgeom.pdf}{\texttt{https://qig.itp.uni-hannover.de/~giulini/papers/Diffgeom.pdf}}\hypersetup{urlcolor=MidnightBlue}]

\bibitem{DeWitt_Book}
{B.~S. DeWitt}, 2003.
\newblock \textit{The Global Approach to Quantum Field Theory, Vol. 1}, volume
  114 of \textit{International Series of Monographs in Physics, 114}.
\newblock
\newblock Clarendon Press, Oxford

\bibitem{DeWitt_QG1}
{B.~S. DeWitt},
\newblock 1967
  \hypersetup{urlcolor=MidnightBlue}\href{http://dx.doi.org/10.1103/PhysRev.160.1113}{\textit{{Quantum
  Theory of Gravity. 1. The Canonical
  Theory}}}\hypersetup{urlcolor=MidnightBlue}.
\newblock Phys. Rev.
\newblock 160 1113

\bibitem{DeWitt:1967ub}
{B.~S. DeWitt},
\newblock 1967
  \hypersetup{urlcolor=MidnightBlue}\href{http://dx.doi.org/10.1103/PhysRev.162.1195}{\textit{{Quantum
  Theory of Gravity. 2. The Manifestly Covariant
  Theory}}}\hypersetup{urlcolor=MidnightBlue}.
\newblock Phys. Rev. 162 1195.


\bibitem{Singer:1981xw}
{I.~M. Singer},
\newblock 1981
  \hypersetup{urlcolor=MidnightBlue}\href{http://dx.doi.org/10.1088/0031-8949/24/5/002}{\textit{{The
  Geometry of the Orbit Space for Nonabelian Gauge Theories.
  (Talk)}}}\hypersetup{urlcolor=MidnightBlue}.
\newblock Phys. Scripta
\newblock 24 817

\bibitem{Singer:1978dk}
{I.~M. Singer},
\newblock 1978
  \hypersetup{urlcolor=MidnightBlue}\href{http://dx.doi.org/10.1007/BF01609471}{\textit{{Some
  Remarks on the Gribov Ambiguity}}}\hypersetup{urlcolor=MidnightBlue}.
\newblock Commun. Math. Phys.
\newblock 60 7

\bibitem{Babelon:1979wd}
{O. Babelon} \protect\BIBand{} {C.~M. Viallet},
\newblock 1979
  \hypersetup{urlcolor=MidnightBlue}\href{http://dx.doi.org/10.1016/0370-2693(79)90589-6}{\textit{{The
  Geometrical Interpretation of the {Faddeev-Popov}
  Determinant}}}\hypersetup{urlcolor=MidnightBlue}.
\newblock Phys. Lett.
\newblock 85B 246

\bibitem{Gribov:1977wm}
{V.~N. Gribov},
\newblock 1978
  \hypersetup{urlcolor=MidnightBlue}\href{http://dx.doi.org/10.1016/0550-3213(78)90175-X}{\textit{{Quantization
  of Nonabelian Gauge Theories}}}\hypersetup{urlcolor=MidnightBlue}.
\newblock Nucl. Phys. B139 1.

\bibitem{AliDieter}
{A. Seraj} \protect\BIBand{} {D. Van~den Bleeken},
\newblock 2017
  \hypersetup{urlcolor=MidnightBlue}\href{http://dx.doi.org/10.1007/JHEP08(2017)127}{\textit{{Strolling
  along gauge theory vacua}}}\hypersetup{urlcolor=MidnightBlue}.
\newblock JHEP
\newblock 08 127\hypersetup{urlcolor=BrickRed} [\eprint{1707.00006}]
  \hypersetup{urlcolor=MidnightBlue}

\bibitem{Ali2}
{A. Seraj},
\newblock 2017
  \hypersetup{urlcolor=MidnightBlue}\href{http://dx.doi.org/10.1007/JHEP06(2017)080}{\textit{{Multipole
  charge conservation and implications on electromagnetic
  radiation}}}\hypersetup{urlcolor=MidnightBlue}.
\newblock JHEP
\newblock 06 080\hypersetup{urlcolor=BrickRed} [\eprint{1610.02870}]
  \hypersetup{urlcolor=MidnightBlue}

\bibitem{strominger2018lectures}
{A. Strominger}, 2018.
\newblock \textit{Lectures on the infrared structure of gravity and gauge
  theory}.
\newblock
\newblock Princeton University Press
\newblock [\eprint{1703.05448}]

\bibitem{BFK}
{D. Burghelea}, {L. Friedlander}, \protect\BIBand{} {T. Kappeler},
\newblock 1992
  \hypersetup{urlcolor=MidnightBlue}\href{http://dx.doi.org/https://doi.org/10.1016/0022-1236(92)90099-5}{\textit{Meyer-vietoris
  type formula for determinants of elliptic differential
  operators}}\hypersetup{urlcolor=MidnightBlue}.
\newblock Journal of Functional Analysis
\newblock 107(1) 34

\bibitem{KirstenBFK}
{K. Kirsten} \protect\BIBand{} {Y. Lee},
\newblock 2015
  \hypersetup{urlcolor=MidnightBlue}\href{http://dx.doi.org/10.1063/1.4936074}{\textit{The
  Burghelea-Friedlander-Kappeler–gluing formula for zeta-determinants on a
  warped product manifold and a product
  manifold}}\hypersetup{urlcolor=MidnightBlue}.
\newblock Journal of Mathematical Physics
\newblock 56(12) 123501\hypersetup{urlcolor=BrickRed}
  \hypersetup{urlcolor=MidnightBlue}

\bibitem{Agarwal}
{A. Agarwal}, {D. Karabali}, \protect\BIBand{} {V.~P. Nair},
\newblock 2017
  \hypersetup{urlcolor=MidnightBlue}\href{http://dx.doi.org/10.1103/PhysRevD.96.125008}{\textit{{Gauge-invariant
  Variables and Entanglement Entropy}}}\hypersetup{urlcolor=MidnightBlue}.
\newblock Phys. Rev.
\newblock D96(12) 125008\hypersetup{urlcolor=BrickRed} [\eprint{1701.00014}]
  \hypersetup{urlcolor=MidnightBlue}

\bibitem{Donnelly:2014fua}
{W. Donnelly} \protect\BIBand{} {A.~C. Wall},
\newblock 2015
  \hypersetup{urlcolor=MidnightBlue}\href{http://dx.doi.org/10.1103/PhysRevLett.114.111603}{\textit{{Entanglement
  entropy of electromagnetic edge modes}}}\hypersetup{urlcolor=MidnightBlue}.
\newblock Phys. Rev. Lett.
\newblock 114(11) 111603\hypersetup{urlcolor=BrickRed} [\eprint{1412.1895}]
  \hypersetup{urlcolor=MidnightBlue}

\bibitem{AronWill}
{W. Donnelly} \protect\BIBand{} {A.~C. Wall},
\newblock 2016
  \hypersetup{urlcolor=MidnightBlue}\href{http://dx.doi.org/10.1103/PhysRevD.94.104053}{\textit{Geometric
  entropy and edge modes of the electromagnetic
  field}}\hypersetup{urlcolor=MidnightBlue}.
\newblock Phys. Rev. D
\newblock 94 104053
\newblock[\eprint{1506.05792}]

\bibitem{Brading2004}
{K. Brading} \protect\BIBand{} {H.~R. Brown},
\newblock 2004
  \hypersetup{urlcolor=MidnightBlue}\href{http://dx.doi.org/10.1093/bjps/55.4.645}{\textit{{Are
  Gauge Symmetry Transformations
  Observable?}}}\hypersetup{urlcolor=MidnightBlue}.
\newblock The British Journal for the Philosophy of Science
\newblock 55(4) 645\hypersetup{urlcolor=BrickRed}
  \hypersetup{urlcolor=MidnightBlue}
  \newblock\hypersetup{urlcolor=BrickRed}
  [\href{http://philsci-archive.pitt.edu/1436/}{\texttt{philsci-archive.pitt.edu/1436/}}]
  \hypersetup{urlcolor=MidnightBlue}

\bibitem{GreavesWallace}
{D. Wallace} \protect\BIBand{} {H. Greaves}, 2014.
\newblock \textit{Empirical Consequences of Symmetries}.
\newblock British Journal for the Philosophy of Science
\newblock 65(1) 59
\newblock[\eprint{1111.4309v1}]

\bibitem{HollandsBMS}
{S. Hollands}, {A. Ishibashi}, \protect\BIBand{} {R.~M. Wald},
\newblock 2017
  \hypersetup{urlcolor=MidnightBlue}\href{http://dx.doi.org/10.1088/1361-6382/aa777a}{\textit{{BMS
  Supertranslations and Memory in Four and Higher
  Dimensions}}}\hypersetup{urlcolor=MidnightBlue}.
\newblock Class. Quant. Grav.
\newblock 34(15) 155005\hypersetup{urlcolor=BrickRed} [\eprint{1612.03290}]
  \hypersetup{urlcolor=MidnightBlue}

\bibitem{francoisthesis}
{J. Fran\c{c}ois}, 2014.
\newblock \href{http://theses.fr/2014AIXM4037}{\textit{Reduction of gauge symmetries: a new geometrical approach}}.
\newblock Ph.D. thesis,
\newblock Aix-Marseille University
  \newblock\hypersetup{urlcolor=BrickRed}
\newblock[\href{http://theses.fr/2014AIXM4037}{\texttt{theses.fr/2014AIXM4037}}]
  \newblock\hypersetup{urlcolor=MidnightBlue}

\bibitem{FrohlichMorchioStrocchi1979}
{J. Fröhlich}, {G. Morchio}, \protect\BIBand{} {F. Strocchi},
\newblock 1979
  \hypersetup{urlcolor=MidnightBlue}\href{http://dx.doi.org/https://doi.org/10.1016/0370-2693(79)90076-5}{\textit{Infrared
  problem and spontaneous breaking of the Lorentz group in
  QED}}\hypersetup{urlcolor=MidnightBlue}.
\newblock Physics Letters B
\newblock 89(1) 61

\bibitem{BalachandranVaidya2013}
{A.~P. Balachandran} \protect\BIBand{} {S. Vaidya},
\newblock 2013
  \hypersetup{urlcolor=MidnightBlue}\href{http://dx.doi.org/10.1140/epjp/i2013-13118-9}{\textit{Spontaneous
  Lorentz violation in gauge theories}}\hypersetup{urlcolor=MidnightBlue}.
\newblock The European Physical Journal Plus
\newblock 128(10) 118
\newblock[\eprint{1302.3406}]

\bibitem{RovelliGauge2013}
{C. Rovelli},
\newblock 2014
  \hypersetup{urlcolor=MidnightBlue}\href{http://dx.doi.org/10.1007/s10701-013-9768-7}{\textit{{Why
  Gauge?}}}\hypersetup{urlcolor=MidnightBlue}
\newblock Found. Phys.
\newblock 44(1) 91\hypersetup{urlcolor=BrickRed} [\eprint{1308.5599}]
  \hypersetup{urlcolor=MidnightBlue}
  
  
  \bibitem{Alinew}
{E.~S. Kutluk}, {A. Seraj}, \protect\BIBand{} {D. Van Den~Bleeken}, 2019.
\newblock \textit{{Strolling along gravitational vacua}}
  \hypersetup{urlcolor=BrickRed} [\eprint{1904.12869}]
  \hypersetup{urlcolor=MidnightBlue}

\end{thebibliography}
%\end{document}

\end{document}